\documentclass[twocolumn]{aastex7}

\newcommand\rguide{$R_{guide}$}
\newcommand\rgc{$R_{GC}~$}

\usepackage{amsmath}
\usepackage{moresize}
\usepackage{soul}
\usepackage{xcolor}

\newcommand{\teff}{T$_{\rm eff}$}
\newcommand{\logg}{$\log(g)$}

\newcommand{\feh}{[Fe/H]}

\newcommand{\ceh}{[Ce/H]}  
\newcommand{\ndh}{[Nd/H]}

\newcommand{\srh}{[Sr/H]}
\newcommand{\yh}{[Y/H]}
\newcommand{\zrh}{[Zr/H]}

\newcommand{\bah}{[Ba/H]}
\newcommand{\lah}{[La/H]}
\newcommand{\euh}{[Eu/H]}
\newcommand{\smh}{[Sm/H]}

\newcommand{\cfe}{[C/Fe]}
\newcommand{\nafe}{[Na/Fe]}

\newcommand{\cefe}{[Ce/Fe]}  
\newcommand{\ndfe}{[Nd/Fe]}

\newcommand{\srfe}{[Sr/Fe]}
\newcommand{\yfe}{[Y/Fe]}
\newcommand{\zrfe}{[Zr/Fe]}
\newcommand{\mofe}{[Mo/Fe]}
\newcommand{\bafe}{[Ba/Fe]}
\newcommand{\lafe}{[La/Fe]}
\newcommand{\eufe}{[Eu/Fe]}
\newcommand{\smfe}{[Sm/Fe]}
\newcommand{\microturb}{$\xi_t$}

\newcommand{\snr}{S/N}

\newcommand{\rproc}{$r-$process}
\newcommand{\sproc}{$s-$process}

\received{\today}

\submitjournal{AJ}

\shorttitle{OCCAM X. Neutron Capture Abundances with Keck/HIRES \& Magellan/MIKE}
\shortauthors{Myers et al.}
\graphicspath{{./}{figures/}}

\begin{document}

\title{The Open Cluster Chemical Abundances and Mapping Survey: X. Neutron Capture \\ Abundances for Distant Open Clusters with Keck/HIRES and Magellan/MIKE\footnote{This paper includes data gathered with the 6.5 meter Magellan Telescopes located at Las Campanas Observatory, Chile, and the 10.0 meter Keck I telescope at the W. M. Keck Observatory, Hawai'i USA.}}

\correspondingauthor{Natalie R.~Myers}

\author[0000-0001-9738-4829]{Natalie R.~Myers}
\affiliation{Department of Physics and Astronomy, Texas Christian University, TCU Box 298840 
Fort Worth, TX 76129, USA }
\email{n.myers@tcu.edu}

\author[0000-0002-0740-8346]{Peter M.~Frinchaboy}
\affiliation{Department of Physics and Astronomy, Texas Christian University, TCU Box 298840 
Fort Worth, TX 76129, USA }
\affiliation{Maunakea Spectroscopic Explorer, Canada-France-Hawaii-Telescope, Kamuela, HI 96743, USA}
\email{p.frinchaboy@tcu.edu}

\author[0000-0001-6533-6179]{Henrique Reggiani}
\affiliation{Gemini Observatory/NSF’s NOIRLab, Casilla 603, La Serena, Chile}
\affiliation{The Observatories of the Carnegie Institution for Science,  Pasadena, CA 91101, USA}
\email{}

\author[0000-0003-3217-5967]{Sarah Loebman}
 \affiliation{Department of Physics, University of California, Merced, 5200 North Lake Road, Merced, CA 95343, USA}
\email{sloebman@ucmerced.edu}

\author[0000-0001-6761-9359]{Catherine Manea}
 \affiliation{Department of Astronomy, The University of Texas at Austin, 2515 Speedway Boulevard, Austin, TX 78712, USA}
 \affiliation{National Science Foundation Astronomy and Astrophysics Postdoctoral Fellow}
\affiliation{Department of Astronomy, Columbia University, 550 West 120th Street, New York, NY 10027, USA}
\email{}

\author[0000-0003-0509-2656]{Matthew Shetrone}
\affiliation{Department of Astronomy, University of California Santa Cruz}
\email{}
 
\author[0000-0001-6476-0576]{Katia Cunha}
\affiliation{Observatório Nacional, Rua General José Cristino, 77, Rio de Janeiro, RJ 20921-400, Brazil}
\affiliation{Steward Observatory, University of Arizona, 933 North Cherry Avenue, Tucson, AZ 85721-0065, USA} 
\email{}

\author[0000-0002-1423-2174]{Keith Hawkins}
 \affiliation{Department of Astronomy, The University of Texas at Austin, 2515 Speedway Boulevard, Austin, TX 78712, USA}
\email{}

\author[0009-0005-0182-7186]{Amaya Sinha}
\affiliation{Department of Physics \& Astronomy, University of Utah, 115 S. 1400 E., Salt Lake City, UT 84112, USA}
\email{} 

\author[0000-0001-6761-9359]{Gail Zasowski}
\affiliation{Department of Physics \& Astronomy, University of Utah, 115 S. 1400 E., Salt Lake City, UT 84112, USA}
\email{}

\author[0009-0000-4049-5851]{John Donor}
\affiliation{Department of Physics and Astronomy, Texas Christian University, TCU Box 298840 
    Fort Worth, TX 76129, USA }
\email{j.donor@tcu.edu}

\author[0000-0003-2602-4302]{Jonah M.~Otto}
\affiliation{Department of Physics and Astronomy, Texas Christian University, TCU Box 298840 Fort Worth, TX 76129, USA }
\email{J.OTTO@tcu.edu}

\author[0009-0008-0081-764X]{Alessa I.~Wiggins}
\affiliation{Department of Physics and Astronomy, Texas Christian University, TCU Box 298840 Fort Worth, TX 76129, USA }
\email{A.IBRAHIM@tcu.edu}

\begin{abstract}
The chemistry of stars provides powerful insight into the history of the Milky Way.  With multiple large-sky spectroscopic surveys that are currently available, using chemistry as a means to study the evolution and history of the Milky Way has flourished. Open clusters have long been used as landmarks to calibrate different age dating methods (e.g., gyrochronology and asteroseismology). In this work, we utilize the SDSS-IV/APOGEE-based Open Cluster Chemical Abundances and Mapping (OCCAM) survey as our foundation for new optical observations; enabling us to characterize neutron-capture abundances for known cluster members. For 56 stars in 18 open clusters, we collected high-resolution (R $>$ 50,000), high-S/N ($>75$ at 5500\AA), spectra from Keck I and Magellan Baade telescopes. With these data, we derive abundances for 23 elements using BACCHUS, including 7 neutron capture abundances not measurable by APOGEE. Finally, we characterize the radial distribution of these neutron-capture elements in the Milky Way.
We find that the second-peak \sproc~and \rproc~abundances exhibit relatively flat gradients in the Milky Way. Although not as distinct, the first-peak \sproc~abundances also have slopes which are shallower than the alpha and iron-peak elements. The differences in the neutron-capture gradients from the lighter elements not just indicates the sources producing these elements are fundamentally different, but that the timescales on which they are produced also differ (especially for the \rproc). Moreover, a metallicity dependence of the AGB stars responsible for producing the heaviest \sproc~abundances may be necessary to consider in Galactic evolution models.  

\end{abstract}



\section{Introduction} \label{sec:intro}

In an inside-out Galactic formation scenario \citep*[e.g.,][]{ELS62}, the Milky Way started as a relatively small amalgamation of primordial gas, stars, and dark matter.  
As new  gas, dust, dark matter, and stars were accreted \citep[e.g.,][]{SZ78}, the Milky Way has gradually grown into the Galaxy we see today. In a broad sense, this implies the Milky Way is composed of regions where there have been many generations of stars formed from recycled gas (the bulge and inner Disk) and other regions that have had relatively few generations of stars (the outer regions of the Galactic Disk and Halo). These two regions are then connected by a continuously decreasing relationship of radius and generations of stars.  This phenomena creates a noticeable difference in the chemical composition of stars which were formed from the inner, more enriched gas, and those formed from the relatively metal-poor outer disk. A metallicity gradient ($\rm \delta [Fe/H]/\delta R$) was initially noticed by \citet{Janes79}, who looked at open clusters as a function of radius and found a steadily decreasing trend in metallicity. Since then, many other open cluster studies have built upon this work to characterize this trend; some extending the relationship to further radii and some focusing on the changes to the gradient based on age groups \citep[e.g.,][]{friel_02, Twarog1997, Carraro2004-knee, yong_2012, bragaglia2008, sestito2008, friel2010, carrera_2011, yong_2012, frinchaboy_13, reddy_16, magrini_2017, occam_p4, Myers_OCCAM, magrini2023-gradients-withESO-clusters, occasoV, Otto-occam}. In addition, other studies utilizing HII regions \citep[e.g.,][]{HII-2017, HII_2022}, planetary nebulae \citep[e.g.,][]{Maciel_2013, Stanghellini_2018}, and classical Cepheids, \citep[e.g.,][]{Yong_2006,daSilva_2023,Nunnari_2026,Trentin_2026} have also aimed to characterize the metallicity distribution of the Disk.

Within the past decade, large samples of open clusters have been used to quantify gradients for {\em other} abundances (e.g., alpha, iron-peak, and odd-Z elements) in the Galactic disk.  Frequently these studies use data available in large-sky surveys to expand their samples of open clusters to homogeneously sample the entire Milky Way \citep{jacobson_16, donor_18, carrera_18, occam_p4, spina_21, Casamiquela_2021_OCchemicaloverlap, Myers_OCCAM, magrini2023-gradients-withESO-clusters, yang2025}.  These studies, while large, are still conducted primarily at medium resolution that limits precision and also with limited measurements of the heavy neutron capture elements needs to explore in details the enrichment patterns within the Milky Way. Beyond metallicity, these other abundances can trace different processes. [Fe/H] itself, as well as the other iron-peak elements, primarily trace Type Ia supernovae enhancement \citep[e.g.,][]{Nomoto_1995}. Type II supernovae enrichment is traced by the alpha elements  \citep[e.g.,][]{Woosley_1995}. Elements beyond the iron-peak, however, are produced through the neutron-capture process \citep{Burbidge-1957}. The neutron capture process happens in regions of high neutron flux, and is often split into two different sub-processes: 
the rapid process (\rproc) and the slow process (\sproc).
The \rproc~occurs within regions of high neutron density (i.e., $10^{20}$ neutrons/cm$^3$), which can happen in events such as exotic core-collapse supernovae (e.g., magnetorotational supernovae or collapsars;  \citealt{Winteler2012, Siegel2019,Reichert2023}) or within compact object-neutron star mergers \citep{Lattimer1974,Symbalisty1982, Nstar_rprocess, Nstar_init}. In general, the \rproc~is necessary for the creation of the heaviest elements on the periodic table (e.g., Eu, Sm, Nd). The \sproc~occurs in regions of lower neutron density ($10^{15}$ neutrons/cm$^{3}$) which can happen inside low-intermediate mass asymptotic giant branch stars (AGB) during He shell burning \citep[e.g.,][]{Cristallo2015}, producing elements such as Ce, Ba, and La. the \sproc~can also occur within the interiors of short-lived massive rotating stars ($M_{\star} < 8M_\odot$), which can produce the lighter \sproc~elements, such as Y, Zr, and Sr \citep[e.g.,][]{Limongi2018}.

Open clusters have long been a tracer for exploring the Galactic disk \citep[e.g.,][]{Janes79,JP94,friel1995-review-OCs-AMR,frinchaboy_13}, as they are loosely bound groups of stars which are co-natal single stellar populations with common initial chemistry \citep[e.g.,][]{lada_lada03-protoclusters,Weidner04,Sinha24}.  In addition, the revolution of large-sky and all-sky surveys, such as ESA \textit{Gaia} \citep{gaia_mission}, have vastly extended the number of known open clusters (i.e., the most recent catalog \citealt{hunt23,cavallo24} has over 5000 open cluster candidates).
These inherent properties, as well as the plethora of available data, make open clusters ideal tools for investigating not just the formation and evolution of the Milky Way, but also the processes which chemically enrich it as a function of time.

This paper is organized as follows. In Section \ref{sec:data} and \ref{sec:analysis}, we describe the targeting, collection, and analysis of the data presented in this work. In Section \ref{sec:results}, we show the Galactic gradients we measure and discuss our findings. Finally, we summarize and conclude in Section \ref{sec:conclusion}.

\begin{figure*}[ht!]
    \epsscale{1.}
 	\plotone{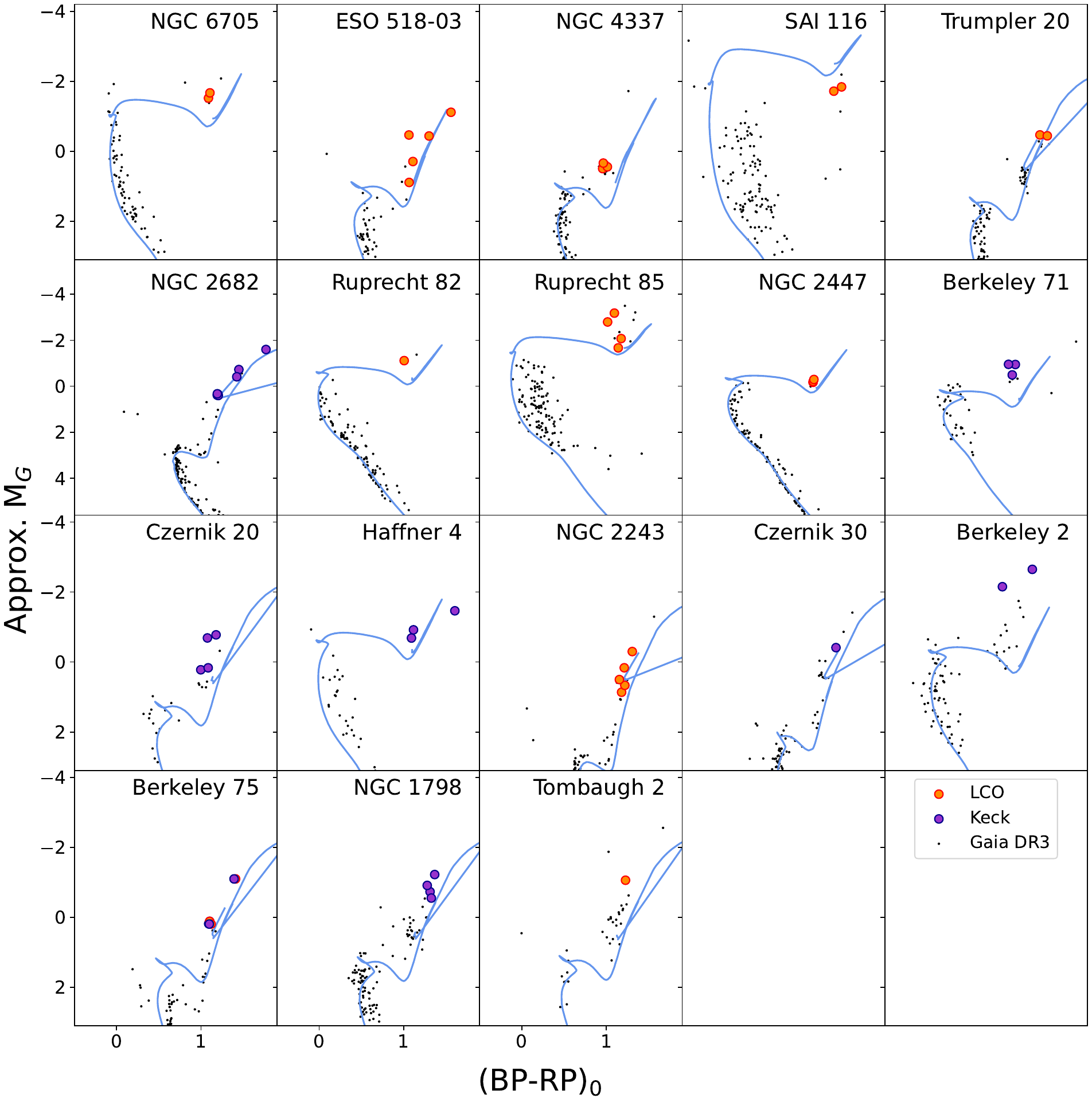}
 	\caption{\small CMDS of each cluster included in this survey, oriented nearest (top row) to farthest (bottom row) from the sun.  The black dots are {\em Gaia}-DR3 members as denoted in \citet{hunt23}, and the colored dots represent stars from this work (orange being the stars taken at LCO, and purple at Keck). All of the points are approximately corrected to absolute magnitudes using the cluster extinction and distance modulus from CG20 (AVNN and DMNN respectively). For clarity, approximate PARSEC isochrones are overplotted in light blue using the CG20 Age and solar metallicity. }
 	\label{fig:CMDS}
\end{figure*}

\section{Observations and Data Reductions} \label{sec:data}

\begin{deluxetable*}{lccrrrrc}[ht!]
 \tabletypesize{\small}
\tablecaption{Cluster Parameters \label{tab:cluster_params}}
\tablehead{
    \colhead{Cluster} &
    \colhead{RA} &
    \colhead{Dec} &
    \colhead{R$_{g}$\tablenotemark{a}} &
    \colhead{R$_{GC}$\tablenotemark{a}} &
    \colhead{Distance\tablenotemark{b}} &
    \colhead{Age\tablenotemark{b}} &
    \colhead{No. of} \\[-2ex]
    \colhead{Name} &
    \colhead{(hms)} &
    \colhead{(dms)} &
    \colhead{(kpc)} &
    \colhead{(kpc)} &
    \colhead{(pc)} &
    \colhead{(Gyr)} &
    \colhead{Stars}
}
\startdata
\hline \hline
\multicolumn{8}{c}{Young Clusters (Age $< 400$ Myr)}\\\hline
SAI 116      & $+11\ 49\ 17.78$ & $-62\ 13\ 33.61$ &   7.47 &   7.71 &  4130 & 0.13 & 2 \\[-0.75ex]
Ruprecht 85  & $+10\ 01\ 26.39$ & $-55\ 06\ 32.50$ &   8.80 &   9.06 &  4803 & 0.20 & 4 \\[-0.75ex]
NGC 6705     & $+18\ 51\ 03.84$ & $-06\ 16\ 19.07$ &   6.40 &   6.40 &  2203 & 0.31 & 2 \\
\hline\multicolumn{8}{c}{Intermediate Clusters ($400 <$ Age $< 800$ Myr)}\\\hline
Ruprecht 82  & $+09\ 45\ 41.04$ & $-54\ 00\ 14.22$ &   8.27 &   9.00 &  2201 & 0.46 & 1 \\[-0.75ex]
Haffner 4    & $+07\ 06\ 02.89$ & $-15\ 00\ 07.23$ &  12.24 &  12.71 &  5051 & 0.46 & 3 \\[-0.75ex]
NGC 2447     & $+07\ 44\ 33.84$ & $-23\ 51\ 10.90$ &   8.83 &   9.20 &  1018 & 0.58 & 3 \\[-0.75ex]
Berkeley 2   & $+00\ 25\ 16.57$ & $+60\ 23\ 31.33$ &  13.16 &  13.06 &  6928 & 0.59 & 2 \\
\hline\multicolumn{8}{c}{Older Clusters (800 Myr $<$ Age $< 2.0$ Gyr)}\\\hline
Berkeley 71  & $+05\ 40\ 55.91$ & $+32\ 16\ 19.30$ &  11.87 &  11.89 &  3598 & 0.87 & 3 \\[-0.75ex]
ESO 518 03   & $+16\ 47\ 02.39$ & $-25\ 48\ 36.05$ &   6.69 &   7.08 &  1631 & 1.41 & 5 \\[-0.75ex]
NGC 4337     & $+12\ 24\ 05.27$ & $-58\ 07\ 30.00$ &   7.39 &   7.49 &  2450 & 1.45 & 4 \\[-0.75ex]
Tombaugh 2   & $+07\ 03\ 05.52$ & $-20\ 49\ 11.98$ &  15.70 &  16.75 &  9316 & 1.62 & 1 \\[-0.75ex]
NGC 1798     & $+05\ 11\ 39.36$ & $+47\ 41\ 27.43$ &  13.20 &  14.11 &  5124 & 1.66 & 4 \\[-0.75ex]
Czernik 20   & $+05\ 20\ 30.96$ & $+39\ 32\ 42.11$ &  11.71 &  12.19 &  3488 & 1.66 & 4 \\[-0.75ex]
Berkeley 75  & $+06\ 49\ 00.48$ & $-23\ 59\ 56.58$ &  14.61 &  13.84 &  8304 & 1.70 & 4 \\[-0.75ex]
Trumpler 20  & $+12\ 39\ 31.70$ & $-60\ 38\ 13.09$ &   7.12 &   7.72 &  3392 & 1.86 & 2 \\
\hline\multicolumn{8}{c}{Very Old Clusters (Age $> 2.0$ Gyr)}\\\hline
Czernik 30   & $+07\ 31\ 11.04$ & $-09\ 56\ 42.20$ &  13.72 &  12.88 &  6647 & 2.88 & 1 \\[-0.75ex]
NGC 2682     & $+08\ 51\ 23.04$ & $+11\ 48\ 50.32$ &   8.90 &   8.97 &   889 & 4.27 & 8 \\[-0.75ex]
NGC 2243     & $+06\ 29\ 34.80$ & $-31\ 16\ 55.22$ &  10.52 &  12.79 &  3719 & 4.37 & 5 \\
\hline\hline
\enddata
\tablenotetext{a}{Radii from \citet{Myers_OCCAM}}
\tablenotetext{b}{Distances and ages from \citet{CG20}.}
\end{deluxetable*}

\subsection{Targeting}
To ensure selection of reliable open cluster members, we utilize the SDSS-IV/APOGEE DR17-based Open Cluster Chemical Abundances and Mapping Survey \citep[OCCAM;][]{Myers_OCCAM} as the basis for our targeting list. OCCAM used \textit{Gaia} EDR3 astrometry \citep{GAIA_EDRthree} with the SDSS-IV APOGEE-2 DR17 {radial velocities (RV)} and metallicities \citep{dr17} to constrain membership for stars in the fields of open clusters. This catalog included 150 unique open clusters, with 95 of them being defined as ``high-quality",{ which is} a designation that was given after visual inspection of the CMD and the distributions in metallicity, radial velocity, and proper motion space. Of those, we further only include those which have age and distance measurements from \citealt[][(hereafter CG20)]{CG20}, which brings the final selection of open clusters to 87. 

Initially, to be observed in this work, clusters had to first have at least $5+$ members which have 
(1) an effective temperature (T$_{eff}$) range between $4000$ and $6000$K (2) a $\log(g) < 3.5$ dex, (3) Gaia G magnitudes brighter than $15$ and BP magnitudes brighter than $14$, and (4) high ($>$50\%) OCCAM membership probabilities in proper motion, radial velocity {\em and} metallicity space provided in the VAC from  \citet{Myers_OCCAM} 
(i.e., the OCCAM probabilities are $>0.5$ in each of \texttt{PM\_PROB}, corresponding to proper motion; \texttt{RV\_PROB}, corresponding to radial velocity; and \texttt{FEH\_PROB}, corresponding to the metallicity). The OCCAM probabilities are computed as described in Figure 2 of \citet[][]{donor_18}, with the modifications described in \citet{occam_p4} and \citet{Myers_OCCAM}.

 With these cuts, we were left with a sample of 11 bright and well-studied clusters, all of which are between $6.4 < R_{GC}\ (kpc) <  12.8$ 
 (NGC 1817, NGC 188, NGC 2204, NGC 2243, NGC 2420, NGC 2682, NGC 6705, NGC 6811, NGC 6819, NGC 752, NGC 7789.).
These clusters
are optimal as calibrators and for comparisons to other surveys. Of these clusters, we include NGC 2682 (M67), NGC 6705, and NGC 2243 in this work.

However, these cuts exclude many reliable open clusters in the OCCAM DR17 catalog which are neither bright enough ($G_{mag}>$ 13), nor populated enough ($N_{OCCAM} < 5$), to make it through the above cuts. To acquire a larger sample of open clusters and probe a larger region of the Galactic disk, we strategically relax these cuts, focusing on clusters which are either observable from the southern hemisphere (Dec $< -10^{\circ}$), or are located in the outer regions of the disk ($R_{GC} > 11$ kpc).   

First, we extend our sample to include stars with $13 < G_{mag} < 15$. This cut adds only two more clusters to the sample; one southern cluster (Trumpler 20) and one outer cluster (NGC 1798, $R_{GC} = 14.1$ kpc).  Finally, we expand our list of clusters to those with at least one OCCAM member.  Any cluster included which only has one OCCAM member, however, \textit{must} have at least two  high quality (\texttt{Prob} $> 0.9$) members in the \citet{CG20} catalog which fall inside the above cuts. By using these extensions, we are able to include thirteen additional OCCAM open clusters, five northern outer clusters and eight southern clusters. 
In total, we include observations for eighteen distinct open clusters including $62$ unique stars.
We present each cluster in Table \ref{tab:cluster_params}, including the distances and ages from \citet{CG20}, and including the Galactocentric and Galactic guiding center radius from \citet{Myers_OCCAM}.
The telescopes and setups used for these observations are described below.

\subsection{Observations} \label{sec:all_obs}
Since the targets for these observations are based on the infrared-based OCCAM survey data, many of the objects we observe are metal-rich giant stars in the Galactic thin disk ($-0.3 \lesssim \rm[Fe/H] \lesssim 0.3$).
For these stars, crowding and blending of different spectral features are significant concerns for measuring the relatively weak neutron capture lines found in the blue regimes of the spectrum.  Therefore, to maximize our ability to measure these lines, we focus on obtaining a high-resolution ($R\sim$ 50,000) spectra for each star. In addition, we also aimed to acquire spectra that have high signal-to-noise ($S/N$).  Specifically, we aimed to obtain a $S/N \ge 75$ at 5500\AA~for each star observed in this work to enable accurate measurement for Barium lines to a 0.1 dex precision. 
Using this criteria, we conducted observations using the using the HIgh Resolution Echelle Spectrograph \citep[HIRES;][]{hires} on the 10-m Keck I telescope and  with the Magellan Inamori Kyocera Echelle \citep[MIKE;][]{LCO-MIKE} spectrograph on the 6.5-m Magellan Baade Telescope \citep{LCO-Magellan}.

We observed 29 open cluster member stars across eight open clusters using the HIRES on the Keck I telescope at the summit of Mauna Kea, in Hawaii, between January and March of 2024. 
For these observations, we acquired high-resolution spectra for distant ($R_{GC} > 11$ kpc, and faint ($13 < V_{mag} < 15.5$) stars in the sample (ignoring NGC 2682, which was used for calibrations). For best results, we use the B5 Decker (slit width of $0.861^"$ giving $R ={\lambda}/{\Delta\lambda} \sim$ 50,000) and configured HIRES to cover a wavelength range of $4240$ \AA\ to $8690$ \AA.

We also observed 38 stars in 12 open clusters with the Magellan Inamori Kyocera Echelle \citep[MIKE;][]{LCO-MIKE} spectrograph on the 6.5-m Magellan Baade Telescope \citep{LCO-Magellan} at Las Campanas Observatory in La Serena, Chile, between August of 2023 and April of 2024. We utilized the 0.5'' slit which results in a resolution $R ={\lambda}/{\Delta\lambda} \sim$ 56,000 in the blue chip and $R \sim44,000$ for the red chip.

\begin{deluxetable*}{lrlrrrccrrc}[t!]
 \tabletypesize{\ssmall}
\tablecaption{Observational Parameters \label{tab:star_obs}}
\tablehead{
    \colhead{Star ID} &
    \colhead{Gaia DR3 ID} &
    \colhead{APOGEE ID} &
    \colhead{G} &
    \colhead{BP} &
    \colhead{RP} &
    \colhead{Obs. Date} &
    \colhead{Exp.} & 
    \colhead{Tot. Exp.} &
    \colhead{SNR}& \\[-2ex]
    \colhead{} &
    \colhead{} &
    \colhead{} &
    \colhead{} &
    \colhead{} &
    \colhead{} &
    \colhead{(mm/dd/yy)} &
    \colhead{(min)} &
    \colhead{(min)} &
    \colhead{(@5853\AA)}
}
\startdata
Berkeley 2 K1\tablenotemark{b}  & 428729206069100928   & 2M00251546$+$6022048 & 13.89 & 15.20 & 12.78 & 01/01/24 & 3 $\times$ 30 &   90  & 69 \\
Berkeley 2 K2\tablenotemark{b}   & 428735940577774336   & 2M00250674$+$6024147 & 14.39 & 15.44 & 13.37 & 01/01/24 & 4 $\times$ 30 &  120 & 101 \\
\hline
Berkeley 71 K1\tablenotemark{b}  & 3448353826632384512  & 2M05404312$+$3217303 & 14.82 & 16.01 & 13.74 & 02/21/24 & 6 $\times$ 30 &  180  & 100 \\
Berkeley 71 K5\tablenotemark{c}  & 3448352898919446400  &                    & 14.36 & 15.48 & 13.26 & 03/19/24 & 5 $\times$ 30 &  150  & 73 \\
Berkeley 71 K6\tablenotemark{b}  & 3448352830199972480  & 2M05405316$+$3215197 & 14.36 & 15.59 & 13.28 & 02/25/24 & 4 $\times$ 30 &  120  & 84 \\
\hline
Berkeley 75 K1\tablenotemark{b} & 2922223081654847232  & 2M06485504$-$2359271 & 13.79 & 14.51 & 12.97 & 01/01/24 & 3 $\times$ 20 &   60  & 99 \\[-0.5ex]
\phantom{Berkeley 75 }L1\tablenotemark{b} & 2922223081654847232  & 2M06485504$-$2359271 & 13.79 & 14.51 & 12.97 & 01/09/24 & 2 $\times$ 67 &  133  & 71\\[0.9ex]
Berkeley 75 K2\tablenotemark{b} & 2922211536782715008  & 2M06491539$-$2359187 & 15.08 & 15.62 & 14.38 & 01/01/24 & 6 $\times$ 30 &  180  & 85\\[-0.5ex]
\phantom{Berkeley 75 }L2\tablenotemark{b} & 2922211536782715008  & 2M06491539$-$2359187 & 15.08 & 15.62 & 14.38 & 01/09/24 & 2 $\times$ 73 &  147  & 31\\[0.9ex]
Berkeley 75 K4\tablenotemark{c} & 2922211090106155392  &                    & 15.00 & 15.54 & 14.31 & 02/27/24 & 4 $\times$ 30 &  120  & 76\\[-0.5ex]
\phantom{Berkeley 75 }L4\tablenotemark{c}   & 2922211090106155392  &                    & 15.00 & 15.54 & 14.31 & 04/03/24 & 3 $\times$160 &  480  & 76\\[0.9ex]
Berkeley 75 K5\tablenotemark{c} & 2922210540350387840  &                    & 15.07 & 15.58 & 14.35 & 02/27/24 & 4 $\times$ 30 &  120  & 81\\
\hline
Czernik 20 K1\tablenotemark{b}  & 187966907248932736   & 2M05203031$+$3928459 & 14.36 & 15.26 & 13.43 & 02/22/24 & 3 $\times$ 30 &   90  & 89\\
Czernik 20 K2\tablenotemark{b}  & 187991787998250112   & 2M05204153$+$3934173 & 14.42 & 15.26 & 13.52 & 02/22/24 & 3 $\times$ 30 &   90  & 87\\
Czernik 20 K6\tablenotemark{c}  & 187991375681421824   &                    & 13.42 & 14.37 & 12.46 & 02/22/24 & 6 $\times$ 20 &  105  & 116 \\
Czernik 20 K7\tablenotemark{c}  & 187991066443744896   &                    & 13.51 & 14.39 & 12.59 & 02/27/24 & 3 $\times$ 17 &   50  & 81 \\
\hline
Czernik 30 K2\tablenotemark{b}  & 3035355578248273536  & 2M07311590$-$0955415 & 14.32 & 15.09 & 13.46 & 02/26/24 & 3 $\times$ 30 &   90  & 81\\
\hline
ESO 518 03 L1\tablenotemark{c}  & 6034635866468406400  &                      & 11.51 & 12.20 & 10.70 & 04/03/24 & 1 $\times$ 22 &   22  & 123\\
ESO 518 03 L2\tablenotemark{b}  & 6046270005147215488  & 2M16464504$-$2558201 & 10.86 & 11.87 &  9.87 & 04/17/23 & 1 $\times$ 13 &   13  & 85 \\
ESO 518 03 L3\tablenotemark{c}  & 6046647996630888832  &                      & 11.53 & 12.38 & 10.63 & 04/04/24 & 1 $\times$ 15 &   15 &  122\tablenotemark{d}\\
ESO 518 03 L4\tablenotemark{b}  & 6046648851319395968  & 2M16470166$-$2547374 & 12.86 & 13.56 & 12.05 & 04/17/23 & 1 $\times$ 60 &   60  & 109\\
ESO 518 03 L5\tablenotemark{b}  & 6046654520671710592  & 2M16472991$-$2544594 & 12.26 & 12.98 & 11.44 & 04/04/24 & 1 $\times$ 45 &   45  & 129\\
\hline
Haffner 4 K1\tablenotemark{b}   & 2936304698871708032  & 2M07060294$-$1459361 & 13.88 & 14.65 & 13.03 & 02/26/24 & 3 $\times$ 20 &   60  &85\\
Haffner 4 K2\tablenotemark{c}   & 2936303496280878720  &                    & 13.11 & 14.20 & 12.07 & 02/26/24 & 3 $\times$ 15 &   45  & 41\\
Haffner 4 K3\tablenotemark{c}   & 2936304698871708672  &                    & 13.65 & 14.39 & 12.74 & 02/26/24 & 3 $\times$ 15 &   45  & 63\\
\hline
NGC 1798 K1\tablenotemark{b}    & 213086354194500224   & 2M05112446$+$4740027 & 13.44 & 14.41 & 12.48 & 01/01/24 & 6 $\times$ 20 &  150  & 173\\
NGC 1798 K2\tablenotemark{b}    & 213087866022954880   & 2M05113666$+$4741482 & 13.75 & 14.66 & 12.82 & 02/25/24 & 3 $\times$ 20 &   60  & 86\\
NGC 1798 K3\tablenotemark{b}    & 212899849533910144   & 2M05114006$+$4739238 & 13.93 & 14.86 & 12.99 & 02/25/24 & 3 $\times$ 20 &   60  & 68\\
NGC 1798 K4\tablenotemark{b}    & 213088003461875072   & 2M05114626$+$4743422 & 14.12 & 15.05 & 13.16 & 03/19/24 & 3 $\times$ 30 &   90  & 74\\
\hline
NGC 2243 L1\tablenotemark{b}    & 2894698045000386816  & 2M06290304$-$3109275 & 13.54 & 14.08 & 12.85 & 02/04/23 & 1 $\times$ 77 &   77  &92\\
NGC 2243 L2\tablenotemark{b}    & 2893947353436037248  & 2M06290934$-$3110325 & 13.73 & 14.25 & 13.06 & 02/05/23 & 2 $\times$104 &  209  & 82\\
NGC 2243 L3\tablenotemark{b}    & 2893936255240622592  & 2M06291101$-$3120394 & 13.03 & 13.56 & 12.35 & 01/09/24 & 2 $\times$ 67 &  133  & 124\\
NGC 2243 L4\tablenotemark{b}    & 2893942714871393664  & 2M06293009$-$3116587 & 12.57 & 13.15 & 11.85 & 01/08/24 & 2 $\times$ 67 &  133  & 125\\
NGC 2243 L5\tablenotemark{b}    & 2893943539505067264  & 2M06294583$-$3115382 & 13.37 & 13.87 & 12.72 & 01/09/24 & 2 $\times$ 70 &  140  & 121\\
\hline
NGC 2447 L1\tablenotemark{b}    & 5615567317447058816  & 2M07441988$-$2352345 &  9.92 & 10.37 &  9.30 & 02/04/23 & 1 $\times$  4 &    4  & 103\\
NGC 2447 L2\tablenotemark{b}    & 5615569138513177472  & 2M07442573$-$2349529 &  9.80 & 10.26 &  9.17 & 02/04/23 & 1 $\times$  3 &    3  & 85\\
NGC 2447 L3\tablenotemark{b}    & 5614818240792756224  & 2M07443366$-$2351422 &  9.88 & 10.34 &  9.26 & 02/04/23 & 1 $\times$  5 &    5  & 134\\
\hline
NGC 2682 K1\tablenotemark{b}    & 604965664269158656   & 2M08493465$+$1151256 &  9.09 &  9.78 &  8.30 & 01/01/24 & 3 $\times$  1 &    3  & 153\\
NGC 2682 K2\tablenotemark{b}    & 598955115237068032   & 2M08495682$+$1141329 &  9.41 & 10.08 &  8.62 & 01/01/24 & 3 $\times$  1 &    3  & 87\\
NGC 2682 K3\tablenotemark{b}    & 604905156769963136   & 2M08521856$+$1144263 & 10.15 & 10.69 &  9.46 & 03/19/24 & 3 $\times$  5 &   15  & 162\\
NGC 2682 K4\tablenotemark{b}    & 604962193935568896   & 2M08501230$+$1151246 &  8.22 &  9.11 &  7.30 & 02/24/24 & 8 $\times$  5 &   26  & 35\\
NGC 2682 K5\tablenotemark{b}    & 604917629355042176   & 2M08512280$+$1148016 & 10.17 & 10.71 &  9.48 & 01/01/24 & 3 $\times$  3 &    9  & 91\\
NGC 2682 K6\tablenotemark{b}    & 604921512005266048   & 2M08512618$+$1153520 & 10.21 & 10.74 &  9.52 & 02/28/24 & 3 $\times$ 20 &   60  & 161\\
NGC 2682 K7\tablenotemark{b}    & 604920202039656064   & 2M08515952$+$1155049 & 10.21 & 10.75 &  9.52 & 03/19/24 & 3 $\times$  4 &   12  & 159\\
NGC 2682 K8\tablenotemark{b}    & 604922164840316672   & 2M08511269$+$1152423 & 10.22 & 10.77 &  9.53 & 03/19/24 & 3 $\times$  3 &    9  & 163\\
\hline
NGC 4337 L1\tablenotemark{b}    & 6071465481608145664  & 2M12235244$-$5806564 & 13.51 & 14.17 & 12.69 & 02/04/23 & 1 $\times$ 85 &   85  & 99\\
NGC 4337 L3\tablenotemark{b}    & 6071462908936965504  & 2M12241575$-$5808502 & 13.35 & 14.03 & 12.55 & 04/17/23 & 1 $\times$ 90 &   90  & 121 \\
NGC 4337 L4\tablenotemark{b}    & 6071464695643381888  & 2M12235665$-$5807252 & 13.46 & 14.17 & 12.64 & 04/04/24 & 3 $\times$120 &  360  & 120\\
NGC 4337 L5\tablenotemark{b}    & 6071466104392670336  & 2M12240586$-$5807152 & 13.40 & 14.09 & 12.59 & 04/04/24 & 3 $\times$110 &  330  & 93\\
\hline
NGC 6705 L1\tablenotemark{b}    & 4252502649521975936  & 2M18511452$-$0616551 & 11.40 & 12.21 & 10.52 & 08/16/22 & 2 $\times$ 30 &   60  & 113\\
NGC 6705 L2\tablenotemark{b}    & 4252499591505213952  & 2M18511571$-$0618146 & 11.25 & 12.07 & 10.37 & 08/15/22 & 2 $\times$ 65 &  130  & 94\\
\hline
Ruprecht 82 L2\tablenotemark{b} & 5309193178382557184  & 2M09454031$-$5359194 & 11.55 & 12.22 & 10.75 & 01/08/24 & 1 $\times$ 30 &   30  & 117\\
\hline
Ruprecht 85 L1\tablenotemark{b} & 5260122318035965696  & 2M10012950$-$5505127 & 13.95 & 15.30 & 12.82 & 02/05/23 & 1 $\times$ 94 &   94  & 66\\
Ruprecht 85 L2\tablenotemark{b} & 5260121974438737536  & 2M10012976$-$5507408 & 14.35 & 15.68 & 13.23 & 02/04/23 & 1 $\times$ 97 &   97  & 54\\
Ruprecht 85 L3\tablenotemark{c} & 5260098506736314496  &                    & 12.84 & 14.13 & 11.73 & 01/08/24 & 2 $\times$ 67 &  133  & 103\\
Ruprecht 85 L4\tablenotemark{c} & 5260098644175115008  &                    & 13.23 & 14.46 & 12.14 & 01/09/24 & 2 $\times$ 67 &  133 & 75 \\
\hline
SAI 116 L1\tablenotemark{b}     & 5334811318302215808  & 2M11491181$-$6214125 & 13.38 & 14.71 & 12.25 & 02/05/23 & 1 $\times$ 60 &   60  & 62\\
SAI 116 L2\tablenotemark{b}     & 5334808397724470912  & 2M11491918$-$6214038 & 13.50 & 14.76 & 12.40 & 02/05/23 & 1 $\times$ 78 &   78  & 67\\
\hline
Tombaugh 2 L1\tablenotemark{b}  & 2929388564578914176  & 2M07030686$-$2048282 & 14.62 & 15.39 & 13.75 & 02/04/23 & 1 $\times$111 &  111  & 35\\
\hline
Trumpler 20 L4\tablenotemark{b} & 6056527001215561728  & 2M12400451$-$6036566 & 13.09 & 14.00 & 12.15 & 04/04/24 & 2 $\times$ 90 &  180  & 74\\
Trumpler 20 L5\tablenotemark{b} & 6056577961058738688  & 2M12385807$-$6030286 & 13.06 & 13.92 & 12.15 & 04/03/24 & 2 $\times$ 60 &  120  & 105\\
\enddata
\tablenotetext{a}{Star ID is of the form: [Cluster name] [observatory, K or L][index], where K indicates the star was observed at the W. M. Keck Observatory, and L indicates Las Campanas Observatory.}\vskip-0.1in
\tablenotetext{b}{Targeting information from \citealt{Myers_OCCAM}}\vskip-0.1in
\tablenotetext{c}{Targeting information from \citealt{CG20}}\vskip-0.1in
\tablenotetext{d}{BACCHUS did not trust the fits to the Ba line used to derive SNR, we instead use the nearby $5855.1$\AA~Fe I Line.}
\end{deluxetable*}

Overall, we observed 56 stars in 18 open clusters, with Berkeley 75 having three stars observed by both telescopes. Information about each observation, including star IDs, Gaia magnitudes, observation dates and exposure times, \snr~as well as the catalog where we pulled the stars properties from can be found in Table \ref{tab:star_obs}. 
In addition, the ESA \textit{Gaia} Color-Magnitude Diagrams (CMDs) for each cluster are shown in Figure \ref{fig:CMDS} where the stars observed in this work are highlighted in either purple or orange. 
For reference, we also include PARSEC isochrones \citep{Costa25-parsec,bressan12-parsec,bressan1993-parsec,girardi2000-parsec,Nguyen22-parsec} calculated with the CG20 age, reddening, distance, and cluster metallicity.

\subsection{Reductions}

\subsubsection{ W. M. Keck I Observatory -- HIRES}

Data obtained with the W. M. Keck I telescope were reduced using the publicly available MAunaKea Echelle Extraction (MAKEE) pipeline.  We then use the subroutines within the \texttt{echelle} package in IRAF\footnote{IRAF is distributed by the National Optical Astronomy Observatory, which is operated by the Association of Universities for Research in Astronomy (AURA) under a cooperative agreement with the National Science Foundation.} and PyRAF\footnote{PyRAF is a product of the Space Telescope Science Institute, which is operated by AURA for NASA.} \citep{Tody1986-IRAF,Tody1993-IRAF,NOIRLAB-IRAF} to obtain a new wavelength solution (\texttt{ecidentify} and \texttt{ecreidentify}), median combine images of like-objects (\texttt{scombine}), and normalize each image (\texttt{continuum}). 
Finally, we obtain the RV corrections for each star by using the \texttt{fxcor} function in the \texttt{rv} package.

\subsubsection{Las Campanas Observatory -- MIKE}
For the data taken at LCO, we use the standard CarPy pipeline to reduce the images. Overall, after combining the red and blue chips, we find a usable wavelength coverage (on our dataset---due to S/N constraints in the blue) of $3800$ \AA\ to $9000$ \AA. Calibrations for data reduction process (i.e., flats, milky flats, and ThAr lamps) were taken in the afternoon prior to the observations. The reductions were conducted using the standard modules of the data reduction package \textit{CarPy}\footnote{\url{http://code.obs.carnegiescience.edu/mike}} \citep{kelson2000,kelson2003,kelson2014}. Through the \textit{CarPy} pipeline, we obtain a wavelength and flux---calibrated spectrum.  
Afterwards, we use  the same IRAF functions as for the Keck data to normalize the spectra, combine them into a single spectrum, and apply the radial velocity correction. 
Once these steps are complete, we are left with 1D, normalized, and RV corrected spectra which can be analyzed in BACCHUS (discussed in \S \ref{sec:analysis}).

\section{Analysis}\label{sec:analysis}
\subsection{Spectral Analysis with BACCHUS}

To analyze our reduced spectra, we use the Brussels Automatic Code for Characterizing High accUracy Spectra \citep[BACCHUS;][]{bacchus_masseron} to re-determine stellar parameters ($T_{eff}$, $\log(g)$, [Fe/H], and microturbulance) and measure detailed elemental abundances. Although a brief introduction to BACCHUS is included below, a more thorough description can be found in \citet{Hawkins15-s2.2}, \citet{Nelson21-loggerrcut-homogeneity}, and \citet{BAWLAS}.

The BACCHUS code uses the radiative transfer code Turbospectrum \citep{turbospec} and 1D-LTE MARCS model atmosphere grids \citep{marcs_models} to create synthetic spectra which are then fit to the data on a line-by-line basis. For each of the desired lines, BACCHUS computes five synthetic spectra for a $\sim20$\AA \ ``window" around the line, varying only the abundance of that line for each. BACCHUS then uses those models to acquire an abundance using the equivalent width, $\chi^{2}$ minimization,  synthesis, and core fitting (Int) methods. These four methods are described in detail in \citep{BAWLAS}, along with the associated strengths and weaknesses of each.

\begin{figure*}[ht!]
    \epsscale{1.1}
 	\plotone{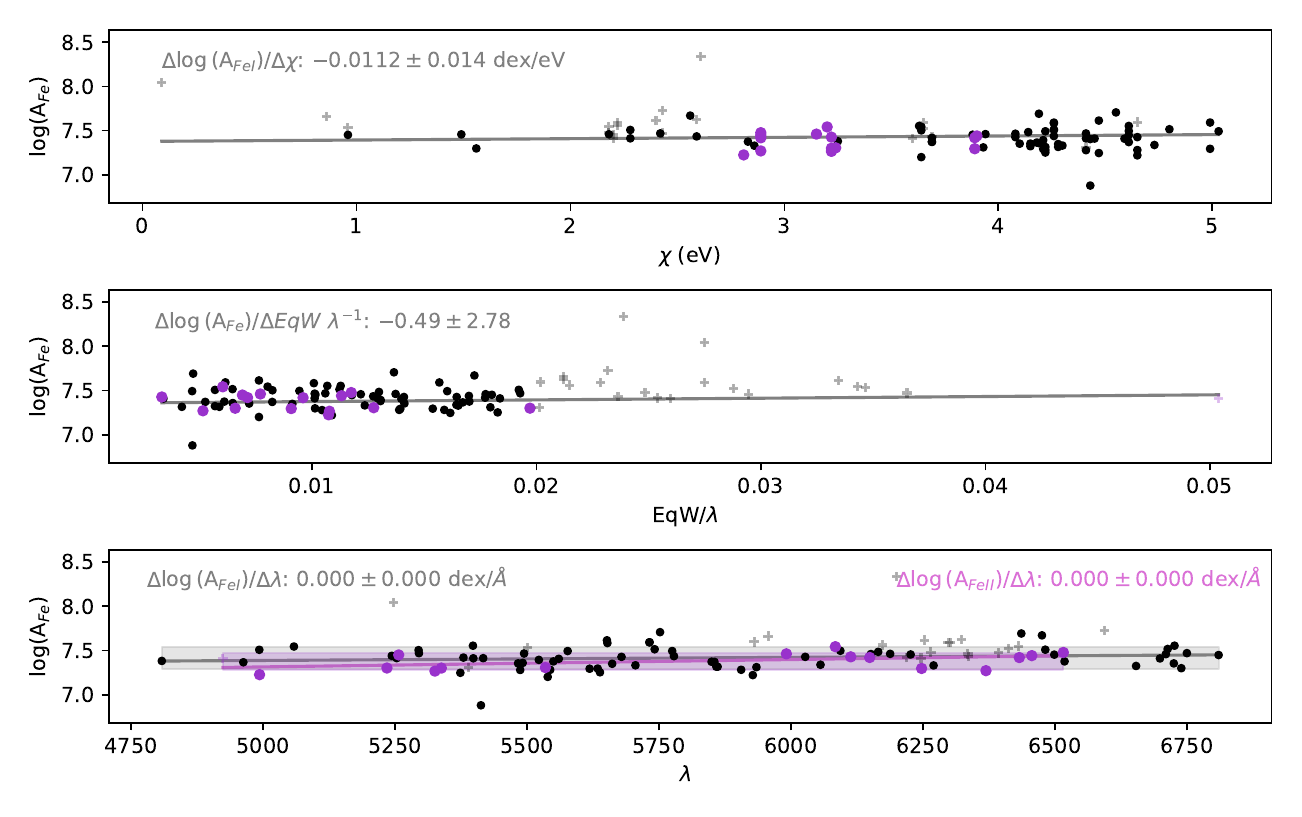}
 	\caption{ \small An example of the excitation and ionization balancing done by BACCHUS to determine the stellar parameters (e.g., \teff, \logg, and \microturb).  The top panel shows abundances from each line as a function of their excitation potential, the middle shows the abundances as a function of their reduced equivalent width, and the bottom shows abundance as a function of wavelength. In all three plots, the abundances derived from Fe~I are black and those derived from Fe~II are shown in purple. The abundances which were derived from the exponential part of the curve of growth (defined to be $EqW/\lambda > 0.02$) are represented by grey $+$ signs. The slopes derived in BACCHUS are shown at the top of each figure in the respective colors, with an example fit provided as well. \teff~is determined when there is a flat trend (within uncertainties) for the $\log(A_{Fe I})$ fit on the top plot. \microturb~converges when the $\log(A_{Fe})$ trend in the middle plot are flat. Finally, \logg~is obtained when the mean abundances from the Fe~I and Fe~II lines match within standard deviation of the measurements (shown as the grey and purple envelopes in the bottom panel).}
 	\label{fig:balancing}
\end{figure*}

\begin{figure}[ht!]
    \epsscale{1.1}
 	\plotone{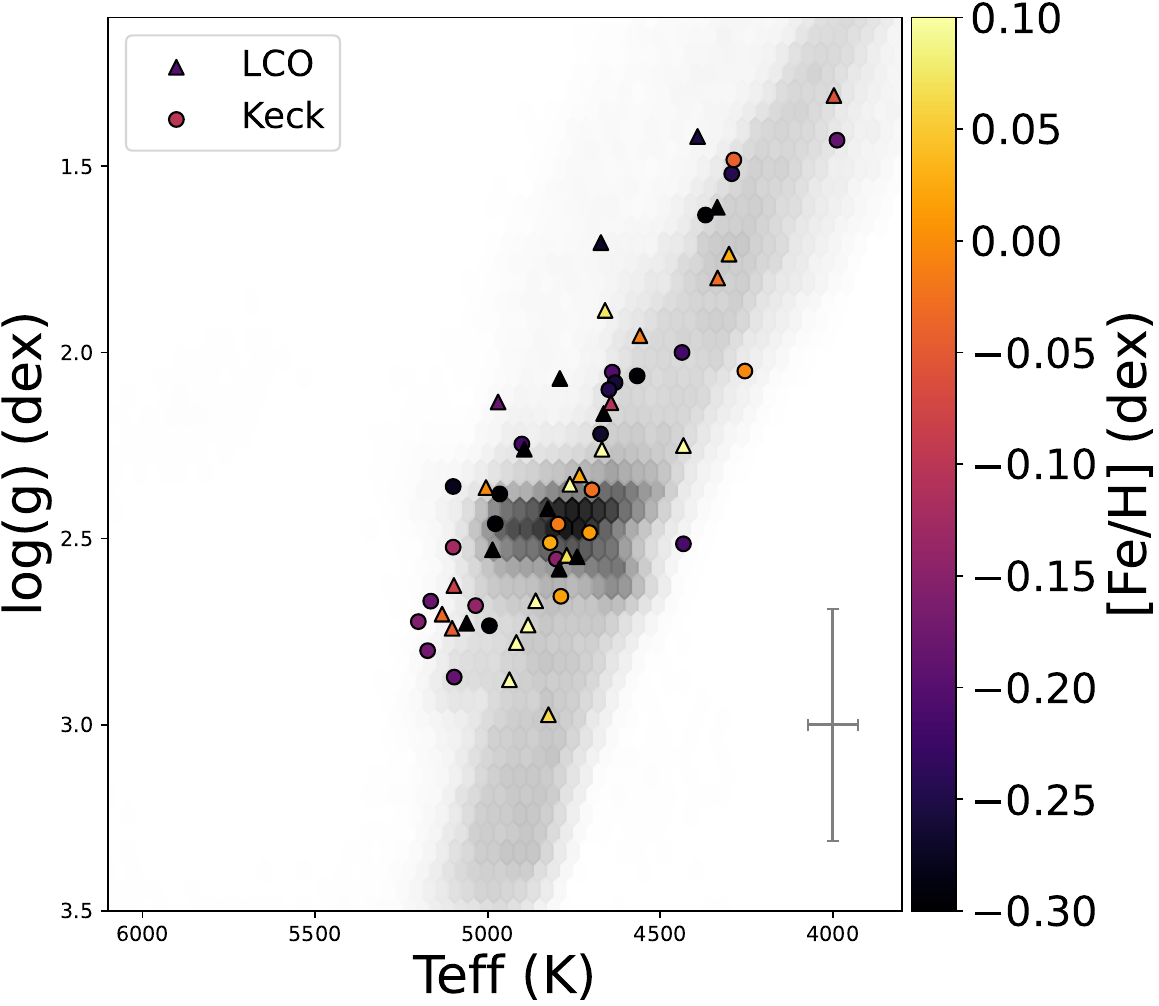}
 	\caption{ \small Kiel diagram for stars presented in this work. The shapes indicate the observatory the star was observed with (Keck or LCO) and the colorbar shows the metallicity of the stars (derived in this work). The APOGEE DR17 Kiel diagram is shown in the background for reference. A representative error bar is shown in the bottom right corner. 
    }
 	\label{fig:kiel}
\end{figure}

To determine the atmospheric parameters of each star, BACCHUS computes the Fe abundances across the spectrum, then uses {a standard }Fe~I and Fe~II excitation and ionization equilibria to determine the T$_{eff}$ and $\log g$ respectively (top panel of Figure \ref{fig:balancing}).  Microturbulance, $\xi_t$, is determined when there is a null trend between the abundance of an element and the reduced equivalent width measured {(middle panel of Figure \ref{fig:balancing})}. Finally, the broadening due to, for example, macroturbulance and stellar rotation ($v \sin(i)$) is estimated with a Gaussian convolution parameter. {If this process converges successfully, the abundance given by each line should be constant at all wavelengths for both Fe~I and Fe~II (bottom panel of Figure \ref{fig:balancing})}. We also include a Kiel diagram in Figure \ref{fig:kiel} with  both the entire APOGEE sample as a hex-grid and our stars with our values for \teff~and \logg.   

\begin{deluxetable*}{cllcc|cllcc|cllcc}[ht!]
 \tabletypesize{\ssmall}
\tablecaption{The Linelist \label{tab:linelist}}
\tablehead{
    \colhead{$\lambda$} &
    \colhead{Elem.} &
    \colhead{Ion} &
    \colhead{E.P } &
    \colhead{$\log gf$} & 
    \colhead{$\lambda$} &
    \colhead{Elem.} &
    \colhead{Ion} &
    \colhead{E.P } &
    \colhead{$\log gf$} & 
    \colhead{$\lambda$} &
    \colhead{Elem.} &
    \colhead{Ion} &
    \colhead{E.P } &
    \colhead{$\log gf$} \\[-2ex]
    \colhead{(\AA)}&
    \colhead{} &
    \colhead{} &
    \colhead{(eV)} &
    \colhead{(dex)} & 
    \colhead{(\AA)}&
    \colhead{} &
    \colhead{} &
    \colhead{(eV)} &
    \colhead{(dex)} & 
        \colhead{(\AA)}&
    \colhead{} &
    \colhead{} &
    \colhead{(eV)} &
    \colhead{(dex)}
}
\startdata
\hline \hline
$4607.331$ & Sr & I & $0.000$ & $+0.28$ & $4741.918$ & Zr & I & $1.775$ & $-0.64$ & $4811.877$ & La & II & $1.847$ & $-2.09$ \\
$4741.918$ & Sr & I & $1.775$ & $-0.41$ & $4811.877$ & Zr & I & $1.847$ & $-0.71$ & $4832.108$ & La & II & $1.798$ & $-0.45$ \\
$4811.877$ & Sr & I & $1.847$ & $+0.19$ & $4832.108$ & Zr & I & $1.798$ & $-1.06$ & $4872.488$ & La & II & $1.798$ & $-1.73$ \\
$4832.108$ & Sr & I & $1.798$ & $-0.11$ & $4872.488$ & Zr & I & $1.798$ & $-1.41$ & $4962.259$ & La & II & $1.847$ & $-3.06$ \\
$4872.488$ & Sr & I & $1.798$ & $-0.06$ & $4962.259$ & Zr & I & $1.847$ & $-1.10$ & $6550.244$ & La & II & $2.690$ & $-2.08$ \\
$4962.259$ & Sr & I & $1.847$ & $+0.20$ & $6550.244$ & Zr & II & $2.690$ & $-0.85$ & $6791.016$ & Ce & II & $1.775$ & $-0.08$ \\
$6550.244$ & Sr & I & $2.690$ & $+0.46$ & $6791.016$ & Zr & II & $1.775$ & $-1.24$ & $6878.310$ & Ce & II & $1.798$ & $-0.08$ \\
$6791.016$ & Sr & I & $1.775$ & $-0.73$ & $6878.310$ & Zr & II & $1.798$ & $-1.16$ & $7070.070$ & Ce & II & $1.847$ & $+0.21$ \\
$6878.310$ & Sr & I & $1.798$ & $-0.24$ & $7070.070$ & Mo & I & $1.847$ & $-1.48$ & $4900.085$ & Ce & II & $1.398$ & $+0.22$ \\
$7070.070$ & Sr & I & $1.847$ & $-0.03$ & $4900.085$ & Mo & I & $1.398$ & $+0.06$ & $6023.406$ & Ce & II & $0.000$ & $+0.07$ \\
$4900.085$ & Y & I & $1.398$ & $-0.33$ & $6023.406$ & Mo & I & $0.000$ & $-0.07$ & $6138.436$ & Ce & II & $0.066$ & $+0.14$ \\
$6023.406$ & Y & I & $0.000$ & $-1.77$ & $6138.436$ & Mo & I & $0.066$ & $-0.34$ & $6222.578$ & Ce & II & $0.000$ & $-0.39$ \\
$6138.436$ & Y & I & $0.066$ & $-1.92$ & $6222.578$ & Mo & I & $0.000$ & $-1.01$ & $6435.004$ & Ce & II & $0.066$ & $-0.81$ \\
$6222.578$ & Y & I & $0.000$ & $-1.45$ & $6435.004$ & Mo & I & $0.066$ & $-1.00$ & $6557.370$ & Ce & II & $0.000$ & $-0.14$ \\
$6435.004$ & Y & I & $0.066$ & $-0.52$ & $6557.370$ & Ba & II & $0.000$ & $-0.91$ & $4854.861$ & Ce & II & $0.992$ & $-0.11$ \\
$6557.370$ & Y & I & $0.000$ & $-2.09$ & $4854.861$ & La & II & $0.992$ & $-0.65$ & $4883.682$ & Ce & II & $1.084$ & $+0.13$ \\
$4854.861$ & Y & II & $0.992$ & $-0.27$ & $4883.682$ & La & II & $1.084$ & $-0.97$ & $4900.119$ & Ce & II & $1.033$ & $-0.40$ \\
$4883.682$ & Y & II & $1.084$ & $+0.19$ & $4900.119$ & La & II & $1.033$ & $-1.08$ & $4982.129$ & Ce & II & $1.033$ & $-1.63$ \\
$4900.119$ & Y & II & $1.033$ & $+0.03$ & $4982.129$ & La & II & $1.033$ & $-0.14$ & $5087.416$ & Ce & II & $1.084$ & $-0.48$ \\
$4982.129$ & Y & II & $1.033$ & $-1.32$ & $5087.416$ & La & II & $1.084$ & $-1.79$ & $5200.406$ & Nd & I & $0.992$ & $-0.28$ \\
$5087.416$ & Y & II & $1.084$ & $-0.16$ & $5200.406$ & La & II & $0.992$ & $-1.24$ & $5205.722$ & Nd & II & $1.033$ & $-0.28$ \\
$5200.406$ & Y & II & $0.992$ & $-0.47$ & $5205.722$ & La & II & $1.033$ & $-1.21$ & $5289.815$ & Nd & II & $1.033$ & $-1.31$ \\
$5205.722$ & Y & II & $1.033$ & $-0.28$ & $5289.815$ & La & II & $1.033$ & $-1.38$ & $5402.774$ & Nd & II & $1.839$ & $-0.48$ \\
$5289.815$ & Y & II & $1.033$ & $-1.68$ & $5402.774$ & La & II & $1.839$ & $-1.05$ & $5544.611$ & Nd & II & $1.738$ & $-0.76$ \\
$5402.774$ & Y & II & $1.839$ & $-0.31$ & $5544.611$ & La & II & $1.738$ & $-1.40$ & $5728.887$ & Nd & II & $1.839$ & $-0.71$ \\
$5544.611$ & Y & II & $1.738$ & $-0.83$ & $5728.887$ & La & II & $1.839$ & $-0.58$ & $6613.732$ & Nd & II & $1.748$ & $-0.90$ \\
$5728.887$ & Y & II & $1.839$ & $-1.15$ & $6613.732$ & La & II & $1.748$ & $-1.16$ & $6795.414$ & Sm & II & $1.738$ & $-0.35$ \\
$6613.732$ & Y & II & $1.748$ & $-0.83$ & $6795.414$ & La & II & $1.738$ & $-1.30$ & $6832.478$ & Sm & II & $1.748$ & $-0.39$ \\
$6795.414$ & Y & II & $1.738$ & $-1.03$ & $6832.478$ & La & II & $1.748$ & $-1.03$ & $7264.161$ & Sm & II & $1.839$ & $-0.65$ \\
$6832.478$ & Y & II & $1.748$ & $-1.94$ & $7264.161$ & La & II & $1.839$ & $-1.36$ & $4784.920$ & Sm & II & $0.687$ & $-1.12$ \\
$7264.161$ & Y & II & $1.839$ & $-1.53$ & $4784.920$ & La & II & $0.687$ & $-1.56$ & $4828.040$ & Sm & II & $0.623$ & $-0.46$ \\
$4784.920$ & Zr & I & $0.687$ & $-0.49$ & $4828.040$ & La & II & $0.623$ & $-1.22$ & $5385.140$ & Eu & II & $0.519$ & $-1.48$ \\
$4828.040$ & Zr & I & $0.623$ & $-0.64$ & $5385.140$ & La & II & $0.519$ & $-2.09$ & $6127.440$ & Eu & II & $0.154$ & $-0.62$ \\
\hline\hline
\enddata
\end{deluxetable*}

\begin{figure*}[ht!]
    \epsscale{1.1}
 	\plotone{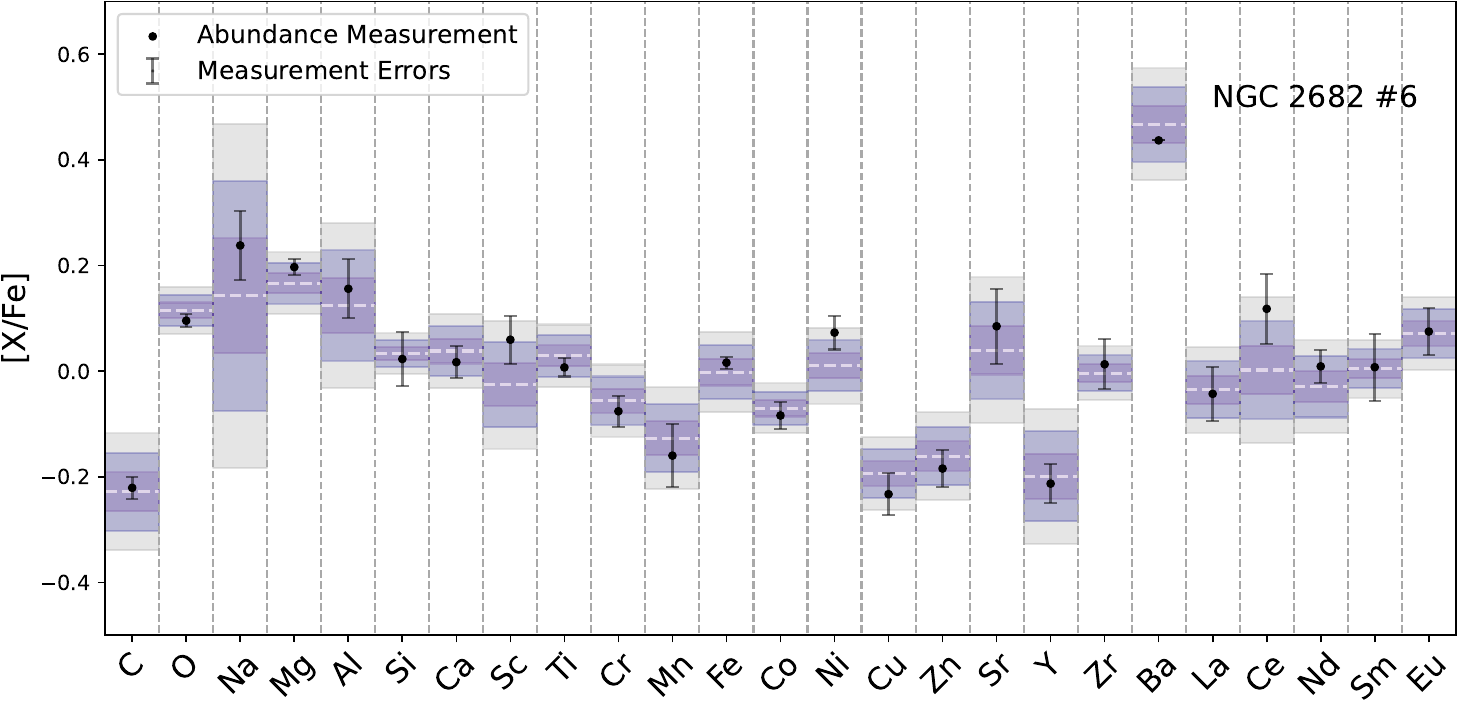}
 	\caption{\small The results of a sensitivity analysis for the star, NGC 2682 \# 6. The final abundances for this star are shown as the black dots with gray error bars. 
    The $1\sigma$, $2\sigma$, and $3\sigma$ standard deviation error envelopes (shown in purple, blue and grey) are computed for each element based off of the eight subsequent BACCHUS runs where \teff, \logg, [Fe/H], or $\xi_t$ are varied within their $1\sigma$ uncertainties (e.g., $T_{eff} \to T_{eff} \pm \sigma_{T_{eff}}$). }
 	\label{fig:sensitivity}
\end{figure*}

Once a stellar atmosphere model is chosen, BACCHUS once again compares this synthetic atmospheric model to the data for each line of the desired element. It synthesizes five different spectra, with four of them perturbed from the calculated ``expected" value (corresponding to the stars' stellar parameters), and calculates abundances using the four distinct methods.   BACCHUS also reports an abundance flag for each method, which can be used to identify suspicious fits or lower/upper limits (see Table 1 in \citealt{BAWLAS} for a description of each flag). In general, we use the $\chi^2$-derived abundance from a line only if none of the four methods were flagged. However, for barium only, we omit the flag for the core fitting method, and rely on the ``chi2" and ``eqw" flags. 
Once all abundances are measured across all lines for a given element, the final star abundance is the mean of the successful measurements minus the corresponding solar abundance from \citet{solabu_A05}.
Finally, the abundance error is given by the standard deviation of each line measured for that element. {It is important to note, that there are stars for which we report abundance errors of $0$ dex due to only one line being measured for that element in that star.  In rare cases where there is only a single line being used to derive an abundance for a star, there may be an abundance error of $0$ dex. }

For this analysis, we use a refined version of the Gaia ESO linelist \citep{GaiaESO,GaiaESO-linelist}. The lines used for the neutron capture abundances in this work are shown in Table \ref{tab:linelist}.
{\color{black} Although not explicitly included in Table \ref{tab:linelist}
 this linelist does include hyperfine structure and isotopic splitting for 
 Sc~\texttt{I},
 V~\texttt{I}, Mn~\texttt{I}, 
 Co~\texttt{I}, Cu~\texttt{I},
 Ba~\texttt{II}, Eu~\texttt{II}, La~\texttt{II}, Pr~\texttt{II}, Nd~\texttt{II}, and Sm~\texttt{I}. 
 To ensure good abundances, we compare the abundances given by each line relative to the cluster mean. If a line consistently yields abundances that are significantly above or below the cluster mean across all stars in which it is measured, we omit that line to avoid introducing bias into our abundance determinations.

 In addition to the overall stellar parameters and abundance measurements, BACCHUS also reports a \snr~about each line it measures. For this work, we use the \snr~computed from the Ba line at $5853$\AA, see Table \ref{tab:star_obs}, except for one star at which this line was insufficiently fitted. In that case, we use a nearby Fe I line ($5855.1$\AA).

 The individual stellar parameters measure by BACCHUS are presented star-by-star in Table \ref{tab:stellar_params} in Appendix \ref{ap:star-by-star}.} 
 We compute open cluster mean abundances by averaging the abundances of the member stars that are presented in \S \ref{sec:results}.

\begin{figure*}[ht!]
    \epsscale{1.1}
 	\plotone{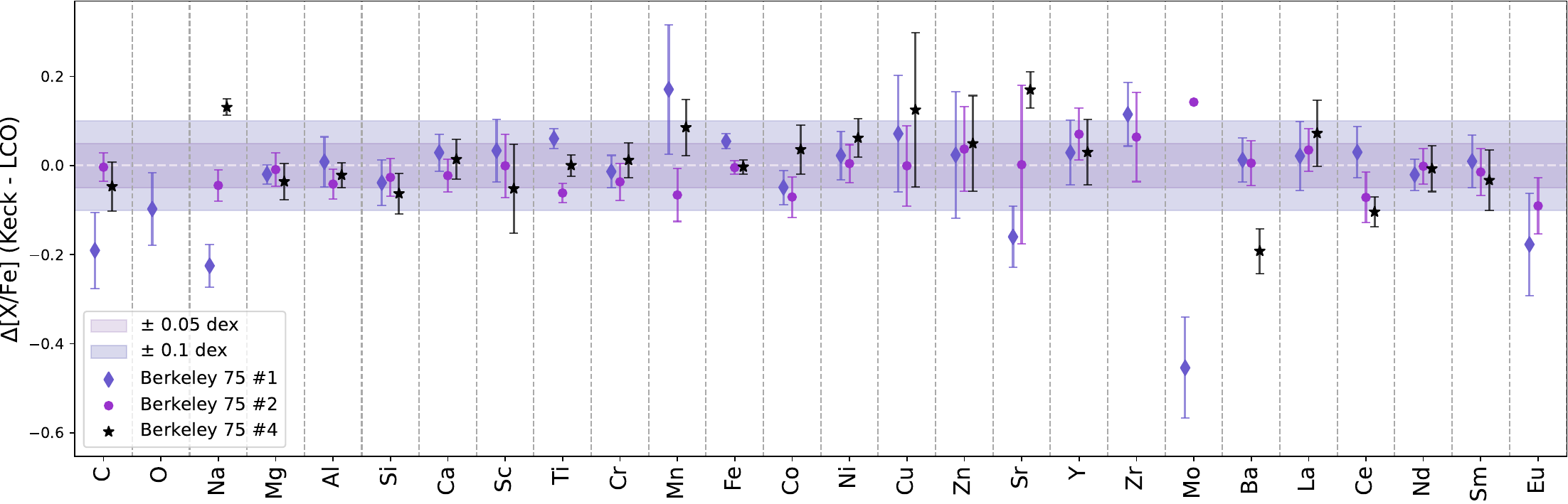}
 	\caption{\small The abundances derived for three stars (blue diamonds for Berkeley 75 \#1, purple dots for Berkeley 75 \#2, and black stars for Berkeley 75 \#4) observed with both the W. M. Keck I Telescope and the LCO Magellan Baade telescope. Here, we compare the abundances for each element as Keck-LCO.  For reference, we include a light purple/blue band at $\Delta$[X/Fe]$ = \pm0.05, \pm 0.1$ dex, respectively. For barium, we add a error of $0.05$dex, corresponding to the standard devation measured from Figure \ref{fig:sensitivity}.
}\label{fig:observatory_comp}
\end{figure*}

\subsection{Parameter and Abundance Verification}

Before continuing with the analysis, we first conduct two main vetting steps. 
\begin{enumerate}
    \item A sensitivity analysis.  
    \item Compare the abundances derived for three stars which were observed by both LCO and Keck. 
\end{enumerate}
\subsubsection{Sensitivity Analysis}
To explore the uncertainties of our abundances, we recalculate the abundances for one of our stars, NGC 2682 \#6, eight different times. With each iteration, we change either the \teff, \logg, \feh, or \microturb \ to its 1$\sigma$ uncertainty value  (e.g., $T_{eff} \to T_{eff} \pm \sigma_{T_{eff}}$) and recalculate the abundances with these stellar parameters. The results of this analysis are shown in Figure \ref{fig:sensitivity}.
Overall, there is good agreement between the original abundance errors and the $1\sigma$ standard deviation. However, there are a few elements where the abundance error is significantly greater than the $1\sigma$ value (i.e., $\rm (1\sigma - [X/Fe]_{err}) > 0.01 $): \cfe, \nafe, \feh, and \bafe. 
By this metric, \cfe~, \nafe, and \feh~underestimate their errors by $\rm0.016\ dex$, $\rm 0.04\ dex$ and $\rm0.014\ dex$, respectively.
The errors for \bafe~are underrepresented because the abundances were only derived from a single line  in the spectrum, leading to a standard deviation of zero. {As such, we add an error floor for \bafe, which is in line with the standard deviation measured by this analysis: $0.05$ dex.}

\subsubsection{Abundance Comparison}

For three stars (Berkeley 75 \# 1, Berkeley 75 \# 2, and Berkeley 75 \# 4) we acquired data at both the Keck and the LCO observatories (details in Appendix \ref{ap:star-by-star}).  To ensure there are no systematics between telescopes, we calculate the difference in the abundance between like-stars and show the results in Figure \ref{fig:observatory_comp}.  
We do see a concerning spread beyond 0.1 dex in certain abundances (C, Na, Mn, Cu, Sr, Mo, Ba, Eu). Of which, the majority (besides Cu, Mo, Na, Ba, Eu) are beyond the uncertainties for the measurements. 
This discrepancy could be due to a myriad of reasons, including differences in: observing conditions, image S/N, \teff, \logg, \feh, and/or \microturb. Below, we explore a few of these parameters in an attempt to understand any underlying systematics which may be affecting the abundance measurements.

Overall, we find good agreement between the stellar parameters derived from each spectra. The largest difference between \teff~measurements is 71 K (Berkeley 75 \#2), the largest difference between \logg~measurements is 0.12 dex (Berkeley 75 \#4), for metallicity the largest difference is 0.06 dex (Berkeley 75 \#1), and the largest difference in \microturb~is $\rm0.12 km\ s^{-1}$ (Berkeley 75 \#1). All of these differences are well within the respective error reported by BACCHUS. Additionally, the stellar parameters are in good agreement with those derived in APOGEE DR17 (Appendix \ref{ap:star-by-star}).

Berkeley 75 \#1 (gray points in Figure \ref{fig:observatory_comp}) shows some of the strongest discrepancies and is also the coolest of the stars. This star resides on the upper region of the RGB as compared to the other two stars (see Figure \ref{fig:CMDS}). Cooler stars have more molecular bands present in their atmosphere, which can make analysis of already weak lines difficult. Additionally, due to the strong differences between \mofe~abundance from different telescopes, and because \mofe~is a highly mixed element which is unreproducible in models \citep[][and references therein]{molero_23}, we omit \mofe~in particular from the analysis beyond this point. For the other elements included here, we continue on with the analysis.

\subsection{Cluster Bulk abundances} \label{sec:bulk_abund}

For each element included in this work, we calculate the respective cluster abundance by taking the mean of each member star's derived value.  The cluster abundance error is derived from the quadrature sum of each abundance error. Our final neutron capture cluster abundances are presented in Table 

In these results, we use the cluster ages presented in CG20, and the guiding center radius (\rguide) computed in \citet{Myers_OCCAM}, see Table \ref{tab:cluster_params}. {The ages in CG20 are computed through an artificial neural network which, when given a CMD that has been binned in magnitude and color space, is able to estimate multiple cluster parameters simultaneously (i.e., age, extinction, and distance).  Although a by-hand isochrone fitting approach may be more accurate, this is a more uniform and efficient method for deriving ages for a large number of clusters. } {\rguide~is the radius of a circular orbit which has the same orbital energy as the corresponding elliptical orbit. That is, for an object in the Milky Way with some Galactocentric radius (\rgc) and an elliptical orbit, \rguide~is the radius of that object if it were on a circular orbit with the same energy.  In \citet{Myers_OCCAM}, this is calculated after first acquiring the Galactocentric coordinates of each cluster, then computing the circular velocity curve using the best-fit Milky Way model described in \citet{Price-Whelan2021}.}  \rguide~is used since it helps mitigate the effects of orbital blurring due to ellipticity in Galactic orbits, which can otherwise cause the clusters to appear further outward or inward from their mean radius.

To quantify the differences in abundance with radius and age, we fit linear trends to our data. We choose a linear fit to match with previous Galactic abundance trend works, both observational \citep[e.g.,][]{Myers_OCCAM, spina_21, magrini2023-gradients-withESO-clusters} and simulation \citep[e.g.,][]{minchev_2018-gradient-evolution, belfiore-2019-model-gradients, johnson-2025-galactic-gradient-model}. 
We use the ensemble sampler from \cite{emcee-python} to determine the maximum likelihood of fit parameters and estimate their uncertainties.

\section{Results and Discussion} \label{sec:results}
With our final derived dataset of abundances, covering 56 unique stars in 18 open clusters across the Galactic disk ($6 \lesssim R_{GC} \lesssim 17$ kpc), we are able to probe 22 different elements total. These elements fall into six nucleosynthetic families: $\alpha$, odd-Z, iron-peak, 1st \& 2nd peak $s-$process, and $r-$process. Although we compute abundances of the lighter elements (e.g., Figures \ref{fig:sensitivity} \& \ref{fig:observatory_comp}) that have also been previously measured by SDSS/APOGEE, here we will primarily focus on measurements of the n-capture abundances and iron-peak elements.

\subsection{Abundance by Radii}\label{sec:radialresults}

With our final sample of open clusters, we explore the variations from solar abundance ratios \citep{solabu_A05} as a function of \rguide. 
We discuss the metallicity gradient in (\S \ref{sec:metallicity-gradient}), 
and the first peak $s-$process elements (Sr, Y, Zr), second peak $s-$process elements (La, Ba, Ce), and $r-$process elements (Eu, Nd, Sm) in \S \ref{sec:ncap-radius}. 

\subsubsection{The Galactic Metallicity Gradient}\label{sec:metallicity-gradient}

\begin{figure}[ht!]
    \epsscale{1.2}
    \plotone{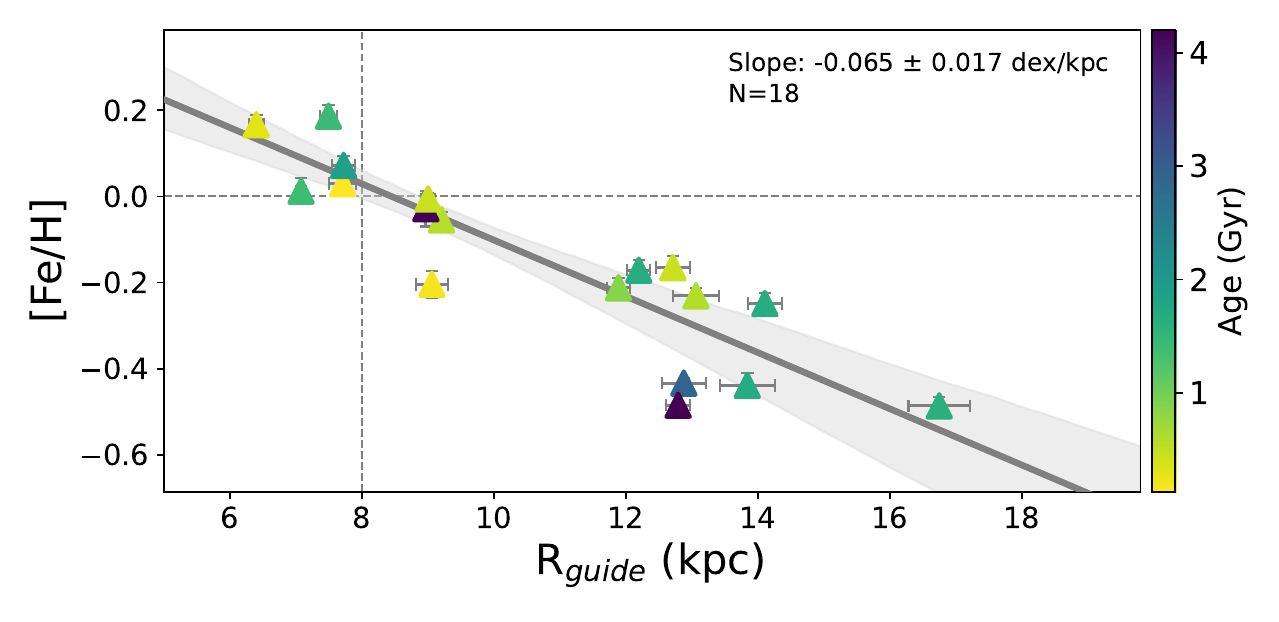}
 	\caption{ \small  Cluster metallicity, represented by \feh, as a function of Galactocentric guiding radius. Each cluster is colored by age in Gyr. The dashed gray lines emphasize the solar abundance (0.0 dex) and solar radius ($\simeq8$ kpc) . Finally, the solid gray line and the light gray envelope are the calculated fit and fit error as calculated by our \textit{emcee} algorithm. The slope for each fit and the number of clusters which have a measurement for the respective element is shown on the plot.  }
 	\label{fig:feh_vs_r}
\end{figure}

\begin{figure}[ht!]
    \epsscale{1.13}
    \plotone{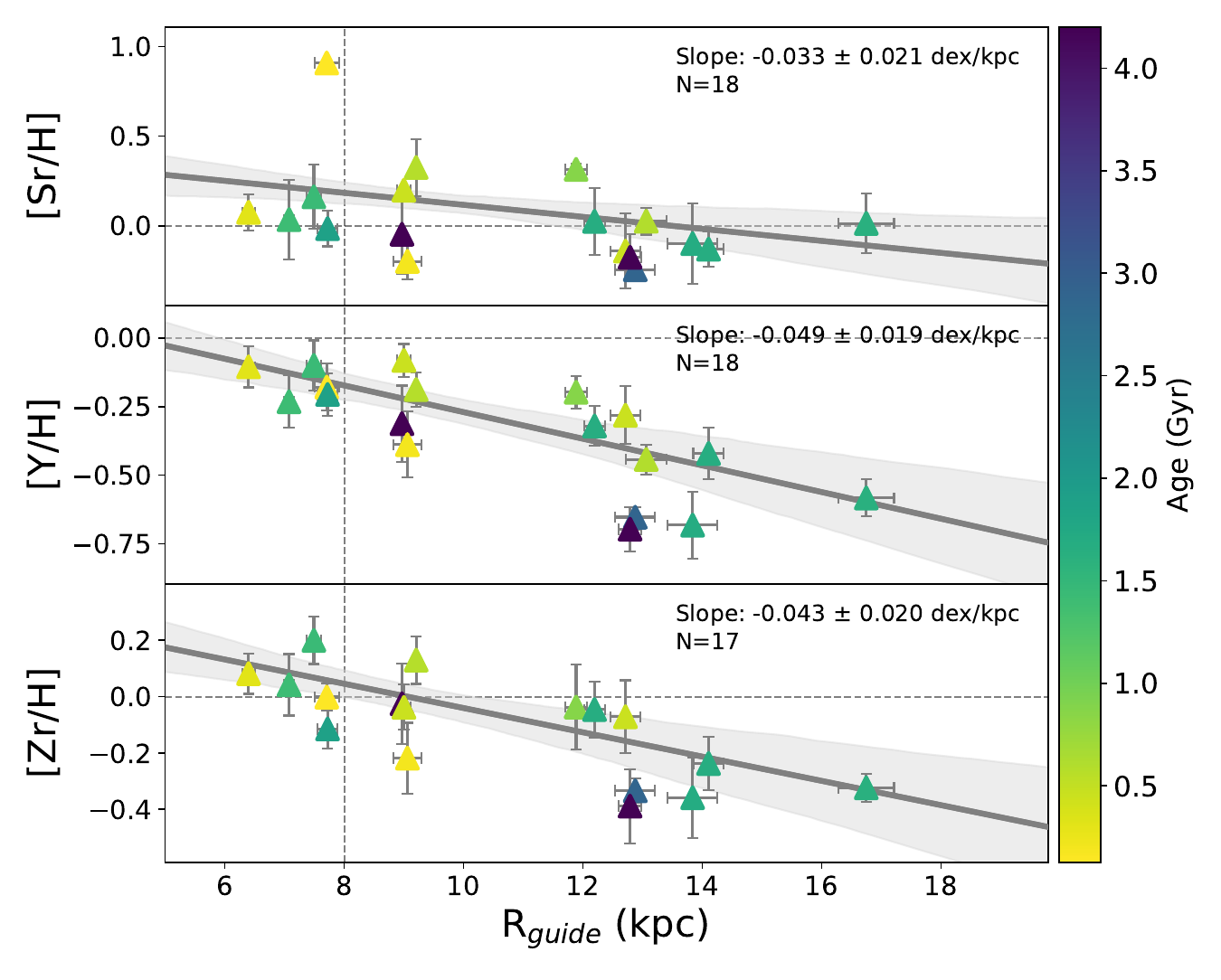}
    \plotone{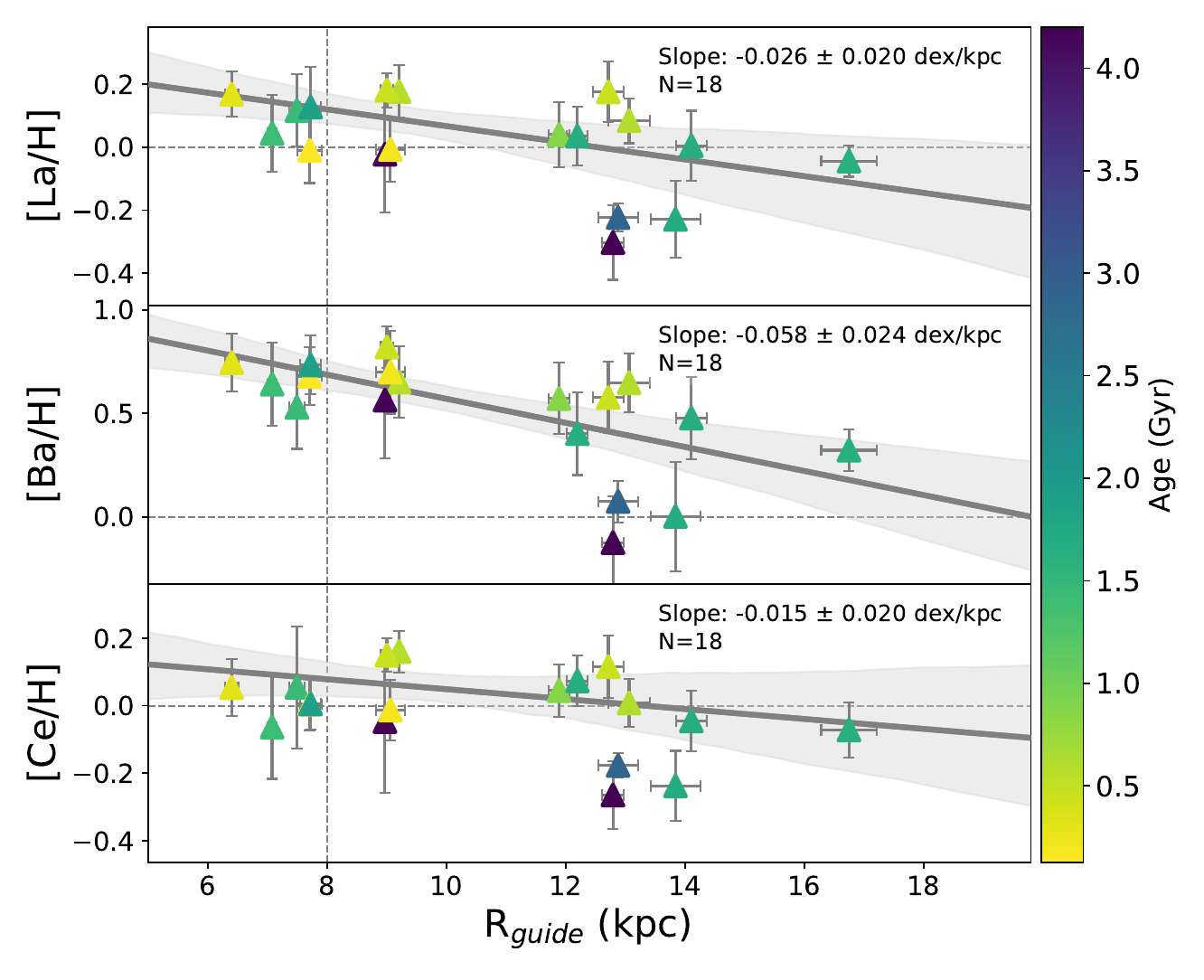}
    \plotone{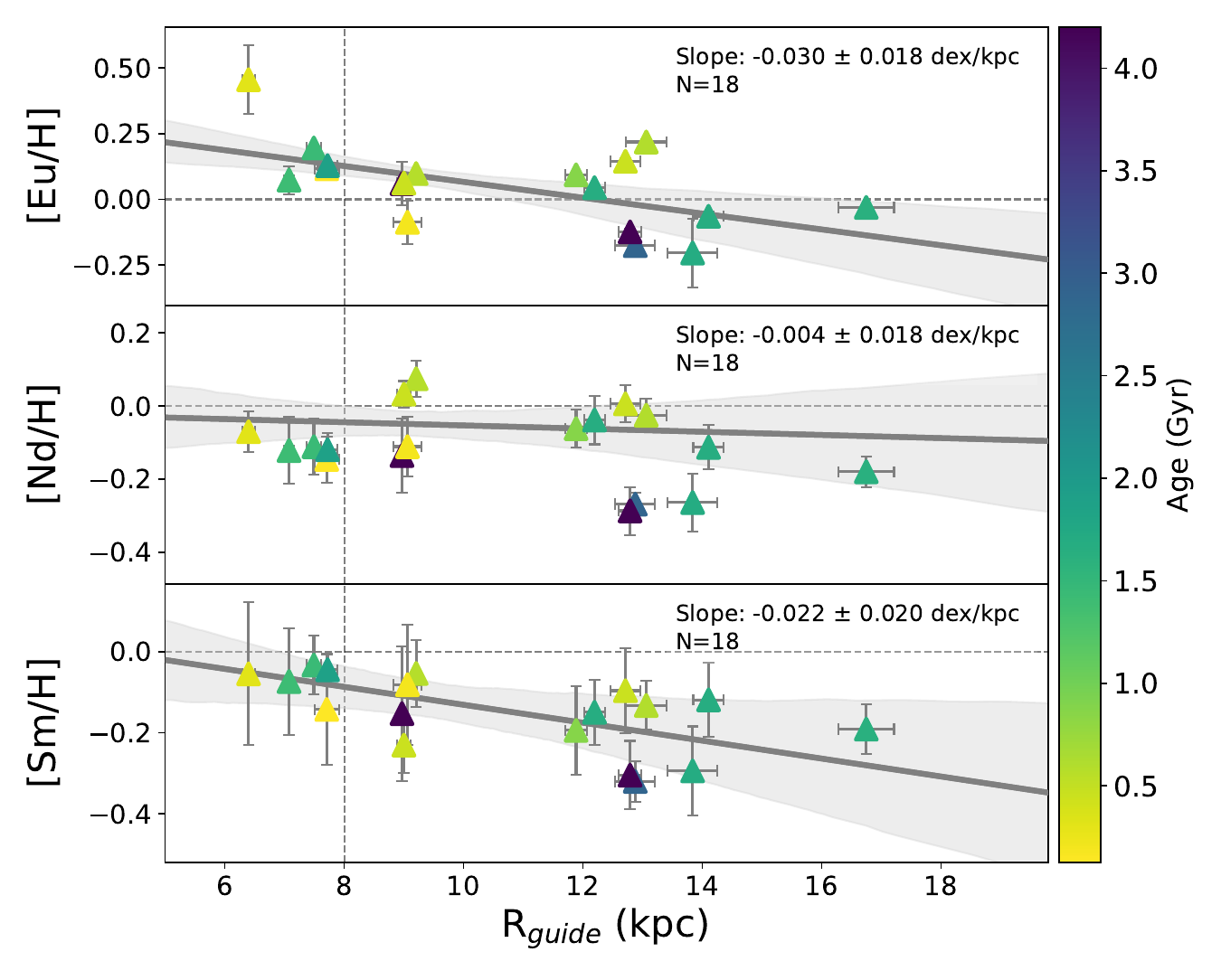}
 	\caption{\small Cluster abundances [X/H] as a function of guiding center radius, \rguide, grouped by abundance family. First-peak $s-$process elements are in the top panel, second-peak $s-$process in the middle, and $r-$process abundances on the bottom. The colors and lines follow the same scheme as in Figure \ref{fig:feh_vs_r}.}
 	\label{fig:allprocess_vs_radius}
\end{figure}

For the metallicity of the clusters, we find a strongly descending slope ($-0.065 \pm 0.017$dex/kpc) as a function of radius. This is consistent with the wealth of previous work using open clusters, whose comparison will be explored in detail in \S \ref{sec:results}. In the literature, it is common to fit this trend with either a single linear fit, or a two-function linear fit and a ``knee" where the two lines intersect, often found around 12 kpc \citep[e.g.,][]{bragaglia2008, sestito2008, friel2010, carrera_2011, yong_2012, frinchaboy_13, reddy_16, magrini_2017, 
occam_p4, Myers_OCCAM, magrini2023-gradients-withESO-clusters, occasoV, Otto-occam}.  Here, we adopt a single linear fit for our sample. We do this in part since we do not include sufficient  clusters at both nearby and large radii to well-constrain the five parameter fit required.

Additionally, \citet{Otto-occam} --- using a substantially larger sample of 158 open clusters from SDSS-V/Milky Way Mapper  \citep[][Johnson, et al., {\em in prep}]{sdss-v} using DR19 \citep{sdss-dr19, sdssv-dr19-ASPCAP} --- found that a linear fit is favored over a bilinear fit based on the Akaike Information Criterion (AIC).

\subsubsection{Neutron Capture Elements}\label{sec:ncap-radius}
We determine abundances for nine different neutron capture elements in this work. Three first peak $s-$process elements, three second peak $s-$process elements, and three elements with significant $r-$process contributions (see Figure \ref{fig:allprocess_vs_radius}). In general, we find these neutron capture elements show significantly shallower slopes than those found for [Fe/H]. 

For the first peak $s-$process abundances (top panel of Figure \ref{fig:allprocess_vs_radius}), we find \srh~and \zrh~to have significantly negative gradients ($-0.033 \pm 0.021$ and $-0.043 \pm 0.020$ dex/kpc, respectively), with \yh~and \zrh~showing the most similarity with the [Fe/H] gradient value ($-0.065 \pm 0.017$dex/kpc). \srh, however, is more shallow, and does not overlap with the \feh~gradient to within the $1\sigma$ errors. 
In addition to the flattening, \yh~also exhibits a much lower abundance throughout all radii than the other two weak $s-$process abundances presented here. Y itself does have an odd atomic number ($\rm Z=39$), whereas Sr and Zr are both even ($\rm Z=38$ and $\rm Z=40$) so the overall difference in abundance could, potentially, be due to the odd-even affect.

Of the second-peak $s-$process elements (middle panel of Figure \ref{fig:allprocess_vs_radius}), \lah~and \ceh~both show shallower trends ($-0.026 \pm 0.020$ and $-0.015 \pm 0.020$ dex/kpc, respectively) than those found in the first-peak \sproc~abundances. \bah, however, shows a significantly steep trend ($-0.058 \pm 0.024$ dex/kpc), similar to those of the first peak \sproc~abundances and the \feh~gradient. 

In the \rproc~elements, we find shallow slopes for \euh~and \smh~which are barely significant within the 1$\sigma$ errors ($-0.030 \pm 0.018$ and $-0.022 \pm 0.020$ dex/kpc, respectively). For \ndh, however, we find a slope which is distinctly flat with radii  ($-0.004 \pm 0.018$ dex/kpc).

\subsubsection{Summary of Galactic Abundance Gradients}

{To summarize these results, we compare the slopes of the neutron-capture abundance gradients to those which are produced by the lighter abundances (Figure \ref{fig:grad_summary}). Of all 13 other elements studied in this work, we have four alpha elements (Mg, Si, Ca, and Ti), two odd-z elements (Na and Sc), and seven iron-peak elements (Cr, Mn, Fe, Co, Ni, Cu, and Zn). 

In general, the iron-peak elements have slopes which closely follow that of the [Fe/H] gradient. The alpha elements, also relatively follow the [Fe/H] gradient, although with a slightly shallower slope. Both of these bulk trends have been shown in the literature \citep[e.g.,][]{Myers_OCCAM, Otto-occam}. 
Cu, Zn, and Sc, however, seem to be slight outliers to this trend. Sc is an odd-Z element, whereas Cu and Zn are both iron-peak elements. The Cu and Zn gradients, in particular, more closely follow the trends of the first-peak \sproc~elements rather than the other iron-peak elements. According to \citet{Arlandini1999} and \citet{Pignatari2010}, Cu and Zn do have some contribution ($\sim1\%$ from Table 2 in \citealt{Arlandini1999}) from the \sproc, which may be why these elements show differing trends. This is also why we choose to omit them in the iron-peak-family envelope in Figure \ref{fig:grad_summary}. 
\begin{rotatetable*}
\noindent\begin{center}
\begin{deluxetable*}{lccccccccccccccccc}
\tabletypesize{\ssmall}
\tablecaption{Cluster Bulk Chemistry \label{tab:full_sample_all_ncap}}
\tablehead{
    \colhead{} && \multicolumn{3}{c}{first peak $s-$process} && \multicolumn{3}{c}{second peak $s-$process}&& \multicolumn{3}{c}{$r-$process}\\[-1ex]\cline{3-5}\cline{7-9}\cline{11-13}
    \colhead{Cluster} &
    \colhead{} &
    \colhead{\bafe}&
    \colhead{\lafe} & 
    \colhead{\cefe} & 
    \colhead{} & 
    \colhead{\srfe} &
    \colhead{\yfe}& 
    \colhead{\zrfe} &
    \colhead{} & 
    \colhead{\ndfe}&
    \colhead{\eufe} & 
    \colhead{\smfe} & 
    \colhead{} \\[-3ex]
    \colhead{Name} &
    \colhead{} &
    \colhead{(dex)} &
    \colhead{(dex)} &
    \colhead{(dex)} &
    \colhead{} &
    \colhead{(dex)} &
    \colhead{(dex)} &
    \colhead{(dex)} &
    \colhead{} &
    \colhead{(dex)} &
    \colhead{(dex)} &
    \colhead{(dex)}
}
\startdata
SAI 116      && $+0.88 \pm +0.01$ & $-0.21 \pm +0.09$ & $-0.03 \pm +0.05$ && $+0.65 \pm +0.01$ & $-0.04 \pm +0.10$ & $-0.02 \pm +0.08$ && $-0.18 \pm +0.06$ & $+0.09 \pm +0.01$ & $-0.17 \pm +0.14$\\
Ruprecht 85  && $+0.01 \pm +0.10$ & $-0.18 \pm +0.12$ & $-0.01 \pm +0.13$ && $+0.90 \pm +0.01$ & $+0.20 \pm +0.10$ & $+0.19 \pm +0.09$ && $+0.09 \pm +0.08$ & $+0.12 \pm +0.08$ & $+0.12 \pm +0.15$\\
NGC 6705     && $-0.09 \pm +0.10$ & $-0.27 \pm +0.07$ & $-0.08 \pm +0.07$ && $+0.58 \pm +0.01$ & $+0.00 \pm +0.07$ & $-0.11 \pm +0.08$ && $-0.24 \pm +0.05$ & $+0.29 \pm +0.13$ & $-0.22 \pm +0.18$\\
Ruprecht 82  && $+0.20 \pm +0.01$ & $-0.08 \pm +0.06$ & $-0.03 \pm +0.08$ && $+0.83 \pm +0.01$ & $+0.19 \pm +0.05$ & $+0.16 \pm +0.05$ && $+0.04 \pm +0.04$ & $+0.07 \pm +0.01$ & $-0.22 \pm +0.07$\\
Haffner 4    && $+0.02 \pm +0.21$ & $-0.12 \pm +0.11$ & $+0.07 \pm +0.13$ && $+0.74 \pm +0.01$ & $+0.34 \pm +0.10$ & $+0.28 \pm +0.09$ && $+0.17 \pm +0.05$ & $+0.36 \pm +0.01$ & $+0.07 \pm +0.11$\\
NGC 2447     && $+0.38 \pm +0.16$ & $-0.13 \pm +0.06$ & $+0.18 \pm +0.08$ && $+0.71 \pm +0.01$ & $+0.23 \pm +0.08$ & $+0.22 \pm +0.06$ && $+0.13 \pm +0.05$ & $+0.15 \pm +0.01$ & $+0.00 \pm +0.08$\\
Berkeley 2   && $+0.26 \pm +0.07$ & $-0.21 \pm +0.05$ & \nodata  && $+0.88 \pm +0.01$ & $+0.31 \pm +0.07$ & $+0.24 \pm +0.07$ && $+0.21 \pm +0.04$ & $+0.45 \pm +0.01$ & $+0.10 \pm +0.06$\\
Berkeley 71  && $+0.49 \pm +0.03$ & $+0.01 \pm +0.06$ & $+0.18 \pm +0.15$ && $+0.78 \pm +0.01$ & $+0.25 \pm +0.10$ & $+0.26 \pm +0.08$ && $+0.15 \pm +0.05$ & $+0.27 \pm +0.01$ & $+0.02 \pm +0.11$\\
ESO 518 03   && $+0.02 \pm +0.22$ & $-0.24 \pm +0.10$ & $+0.03 \pm +0.11$ && $+0.62 \pm +0.01$ & $+0.03 \pm +0.12$ & $-0.08 \pm +0.15$ && $-0.13 \pm +0.09$ & $+0.06 \pm +0.05$ & $-0.09 \pm +0.13$\\
NGC 4337     && $-0.02 \pm +0.18$ & $-0.29 \pm +0.09$ & $+0.01 \pm +0.08$ && $+0.34 \pm +0.01$ & $-0.07 \pm +0.11$ & $-0.13 \pm +0.18$ && $-0.30 \pm +0.08$ & $+0.01 \pm +0.01$ & $-0.21 \pm +0.07$\\
Tombaugh 2   && $+0.50 \pm +0.16$ & $-0.10 \pm +0.07$ & $+0.16 \pm +0.05$ && $+0.76 \pm +0.01$ & $+0.44 \pm +0.05$ & $+0.41 \pm +0.08$ && $+0.31 \pm +0.04$ & $+0.46 \pm +0.01$ & $+0.30 \pm +0.06$\\
NGC 1798     && $+0.12 \pm +0.10$ & $-0.17 \pm +0.09$ & $+0.01 \pm +0.10$ && $+0.73 \pm +0.01$ & $+0.25 \pm +0.11$ & $+0.20 \pm +0.09$ && $+0.14 \pm +0.06$ & $+0.18 \pm +0.02$ & $+0.13 \pm +0.09$\\
Czernik 20   && $+0.20 \pm +0.19$ & $-0.15 \pm +0.07$ & $+0.13 \pm +0.10$ && $+0.57 \pm +0.01$ & $+0.21 \pm +0.09$ & $+0.25 \pm +0.08$ && $+0.13 \pm +0.07$ & $+0.19 \pm +0.01$ & $+0.02 \pm +0.08$\\
Berkeley 75  && $+0.34 \pm +0.22$ & $-0.24 \pm +0.12$ & $+0.08 \pm +0.14$ && $+0.44 \pm +0.01$ & $+0.21 \pm +0.12$ & $+0.20 \pm +0.10$ && $+0.17 \pm +0.08$ & $+0.24 \pm +0.13$ & $+0.14 \pm +0.11$\\
Trumpler 20  && $-0.09 \pm +0.10$ & $-0.28 \pm +0.08$ & $-0.19 \pm +0.07$ && $+0.66 \pm +0.01$ & $+0.06 \pm +0.13$ & $-0.07 \pm +0.08$ && $-0.19 \pm +0.05$ & $+0.06 \pm +0.01$ & $-0.12 \pm +0.04$\\
Czernik 30   && $+0.19 \pm +0.02$ & $-0.22 \pm +0.04$ & $+0.10 \pm +0.04$ && $+0.51 \pm +0.01$ & $+0.21 \pm +0.04$ & $+0.26 \pm +0.04$ && $+0.17 \pm +0.03$ & $+0.26 \pm +0.01$ & $+0.11 \pm +0.05$\\
NGC 2682     && $-0.02 \pm +0.22$ & $-0.28 \pm +0.14$ & $+0.00 \pm +0.14$ && $+0.59 \pm +0.01$ & $+0.01 \pm +0.18$ & $-0.02 \pm +0.21$ && $-0.11 \pm +0.10$ & $+0.09 \pm +0.08$ & $-0.12 \pm +0.17$\\
NGC 2243     && $+0.31 \pm +0.13$ & $-0.21 \pm +0.08$ & $+0.10 \pm +0.13$ && $+0.36 \pm +0.01$ & $+0.18 \pm +0.12$ & $+0.22 \pm +0.10$ && $+0.20 \pm +0.07$ & $+0.36 \pm +0.01$ & $+0.18 \pm +0.08$\\
\enddata
\end{deluxetable*}
\end{center}
    
\end{rotatetable*}

\begin{figure*}[ht!]
    \epsscale{1.1}
 	\plotone{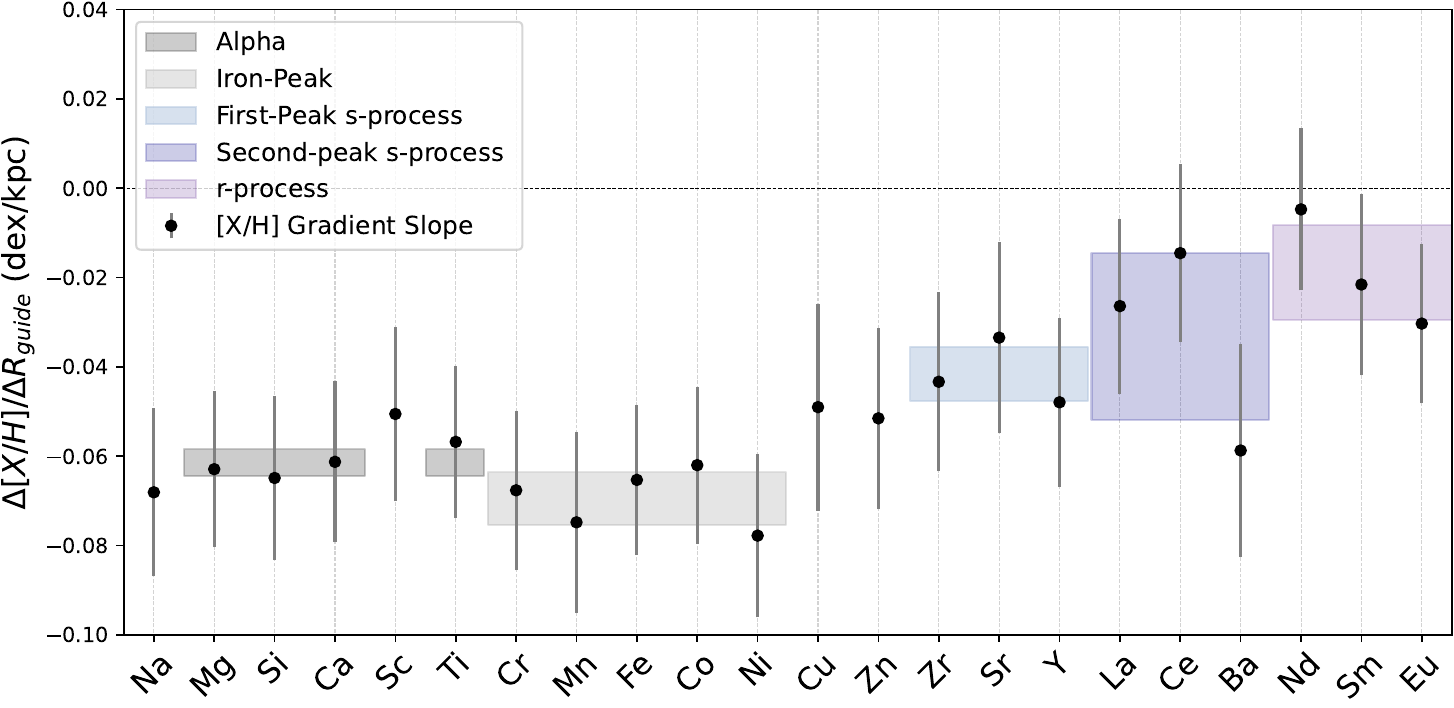}
 	\caption{\small A summary of all 22 measured abundance ([X/H]) gradients in this work, shown as black dots.  For each abundance family, we include a shaded region corresponding to the mean $\pm$ standard deviation of the families slope values. The $s-$process abundances are ordered by ascending \sproc~contribution until Sm and Eu, in which case, they are dominated by the \rproc. }
 	\label{fig:grad_summary}
\end{figure*}

As an odd-Z element, the nucleosynthetic processes which produce Sc primarily occurs within core collapse supernovae similar to the other odd-Z and alpha elements. However, to recreate the observed abundances in Sc, models have required contributions from hypernovae jet effects and the neutrino-process \citep[e.g., see][and references therein]{Kobayashi_2011_F,Sneden_2016, Kobayashi20-C2U}, which may contribute to the shallower gradient seen here. 
These same processes can also produce Ti, V, and Co, two of which are measured in this work, however, the two other abundances measured in this work do not show the same drastic enhancement in the slope as Sc does. 

The neutron-capture gradients, stand out as compared to these lighter-element slopes. Specifically, elements which are enriched through the second-peak \sproc~(low-intermediate AGB), \rproc~(compact object mergers), and  to a lesser degree the first-peak \sproc~(massive rotating stars) show significantly shallower slopes than the other elements. In particular, these slopes seem to get shallower with \sproc~contribution. That is, for each of these neutron-capture elements, the relative contributions from the \sproc~and \rproc~varies. According to Table 1 in \citet{Bisterzo2014}, the relative contribution of the \sproc~for the first-peak abundances average $\sim69\%$ and for the second-peak, $\sim81\%$. The relative \sproc~contribution for \euh~and \smh~is extremely low ($\sim6\%$ and $31\%$, respectively), since these two elements are dominated by the \rproc. \ndh~is the inflection point between elements dominated by the \sproc and those dominated by the \rproc. The \ndh~gradient shows the flattest slope and has a relative \sproc~contribution of $57.5\%$, where the other $\sim43\%$ is produced through the \rproc.

{The \bah~gradient is an obvious outlier to this trend, with a gradient more consistent with the alpha elements than with the other two second-peak \sproc~elements. Ba itself is a difficult element to measure in metal-rich stars, as the lines are relatively strong, {and thus the abundances have to be derived from the non-linear portion of the Curve of Growth. }
Ba is also subject to distinct non-Local Thermodynamic Equilibrium (NLTE) effects, which strengthen as the metallicity of the stars increase \citep{Mashonkina1999-Ba}. Although computing NLTE abundances is beyond the scope of this work, those corrections could act to enshallow the Ba gradient and make it more consistent with the gradients of Ce and La.}
}

\subsection{On Galactic Chemical Evolution}

In the context of inside-out formation, where the distribution of elements like Iron is a consequence of the star formation history (SFH) and delay time distribution (DTD) for different enrichment mechanisms, a flat gradient is relatively unexplained. 

When considering the differences between gradients of the younger populations and  older populations of open clusters, it is evident that the metallicity gradient does not remain stagnant through time \citep[e.g.,][]{occam_p4, Myers_OCCAM,Otto-occam}. Radial migration is expected to play a role in this phenomena as interactions with spiral arms and the bar over time can affect the orbits of the stars in the Milky Way  \citep{sellwood+binney_2002}.  In simulations, \citet{wiggins_2025} found an analog for the old open cluster Berkeley 20, and was able to characterize the interactions which placed it at the high $Z_{max}$ and location we see today. \citet{VV_2023} also compared a sample of 41 open clusters to field stars to characterize the differences in orbital parameters between these two populations. They found that although clusters with ages less than 2--3 Gyr have relatively circular orbits, the older open clusters do have orbits of significantly higher eccentricities; an expected consequence of these orbital perturbations. 

However, all the open clusters in this sample besides two, have ages less than 3 Gyr. Thus any perturbations in the Radial gradient caused by migration should be small. In addition to this, any smoothing of the gradients via radial mixing should affect the different abundance families similarly -- i.e., the alpha gradients cannot be smoothed without also further smoothing the neutron-capture gradients. Therefore, this cannot be the sole reason for the flattening of the neutron-capture gradients. 

Another possible explanation could lie in the metallicity dependence of yields for the neutron-capture process. For AGB stars of lower metallicity, the ratio of free neutrons to free seed nuclei is larger than for those of higher metallicity \citep[e.g.,][]{Cristallo_2011, Bisterzo2014, Prantzos_2020}. When the ratio of free neutrons to seeds is large, more heavy \sproc~abundances (e.g., Ba, Ce, La, and Pb not studied in this work) are produced. When the ratio of free neutrons is small, \sproc~production is dominated by the lighter \sproc~abundances (e.g., Y, Sr, and Zr). This phenomena would work to flatten the gradients, as the stars on more metal-poor regions of the Galactic Disk will have a higher ratio of the second-peak \sproc~abundances. Moreover, once stars begin to become more metal rich, as they are in the solar neighborhood and inner Disk, the relative contribution to the second-peak \sproc~will diminish. This phenomena also explains why the first-peak \sproc~elements more closely follow the lighter element gradients, as their production is limited until more metal-rich stars form. 
The \rproc~production mechanisms, however, are less understood. The only confirmed production sight are neutron-neutron star mergers, which have, to the authors knowledge, no confirmed metallicity dependence in their production yields. More work will be needed to better understand the distribution of these elements. 

\section{Conclusion }\label{sec:conclusion}
In this work, we acquired new observations for 56 stars in 18  open clusters using the high-resolution instruments at the Keck and Las Campanas Observatories.
For each star in our sample, we use IRAF and BACCHUS to derive a suite of 22 different abundances, including: alpha elements (O, Mg, Si, Ca), odd-Z elements (Na, Sc), iron-peak elements (Ti, Cr, Mn, Co, Ni, Cu, Zn), first peak $s-$process elements (Sr, Y, Zr), second peak $s-$process elements (La, Ba, Ce), and $r-$process elements (Eu, Nd, Sm). We then use this sample to probe the Galactic abundance gradients for each nucleosynthetic family in the Milky Way. 

We find a significant difference between the gradients of the lighter alpha and iron-peak families as compared to the neutron-capture (see Figure \ref{fig:grad_summary}). Whereas the iron-peak and alpha families exhibit very similar, relatively steep ($\sim-0.06$dex/kpc) slopes; the second-peak \sproc~and \rproc~abundances show a relatively flat distribution in the Milky Way ($\sim -0.02$ dex/kpc). This trend seems to correlate directly with the relative contribution of \sproc~per element \citep[e.g.,][]{Bisterzo2014}. 

In particular, this indicates:
\begin{itemize}
    \item The nucleosynthetic processes which produce the neutron-capture abundances are distinct from the Type Ia and Type II enrichment sources for the iron-peak and alpha elements.
    
    \item Cu and Zn, although often considered just iron-peak elements, have some significant contribution from the 
    \sproc, which is visible in their gradients. 
    
    \item It has been shown that \sproc~production in AGB stars depends on metallicity.  Considering that the production of the heaviest (second-peak) \sproc~abundances primarily occurs at lower metallicity, and this production rate lowers at higher metallicity. The shallowness of the second-peak \sproc~gradients may be explained purely from this dependence. 
    
    \item Although less prominent, AGB metallicity dependence can also explain the slope of the lighter (first-peak) \sproc~gradients. As these elements are primarily produced at higher metallicity, their gradients will more closely match those of the Type Ia and Type II byproducts.
    
    \item We find the \rproc~abundances closely follow that of the second-peak \sproc~abundances. However, the only confirmed source of \rproc~elements are neutron star mergers. Whether the yields of these processes is also dependent on metallicity is unclear. 
    
\end{itemize}

\section{acknowledgments}
NRM and PMF would like to acknowledge Erika Holmbeck for useful discussions regarding the MIKE observations and survey planning.

Some of the data presented herein were obtained at Keck Observatory, which is a private 501(c)3 non-profit organization operated as a scientific partnership among the California Institute of Technology, the University of California, and the National Aeronautics and Space Administration. The Observatory was made possible by the generous financial support of the W. M. Keck Foundation.
We also wish to extend our special thanks to those of Hawaiian ancestry on whose sacred mountain of Mauna Kea we are privileged to be guests. Without their generous hospitality, the Keck observations presented herein would not have been possible.
This research has made use of the Keck Observatory Archive (KOA), which is operated by the W. M. Keck Observatory and the NASA Exoplanet Science Institute (NExScI), under contract with the National Aeronautics and Space Administration.

NRM, JD, and PMF, acknowledge support for this research from the National Science Foundation (AST-1715662 \& AST-2206541). 
SRL acknowledges support from NSF grant AST-2109234 \& AST-2511388 and HST grant AR-16624 from STScI.
KC acknowledge support for this research from the National Science Foundation (AST-2206543)
PMF and SRL acknowledge some of this work was performed at the Aspen Center for Physics, which is supported by National Science Foundation grant PHY-1607611.

Funding for the Sloan Digital Sky Survey IV has been provided by the Alfred P. Sloan Foundation, the U.S. Department of Energy Office of Science, and the Participating Institutions. 

SDSS-IV acknowledges support and resources from the Center for High Performance Computing  at the University of Utah. The SDSS website is www.sdss.org.

SDSS-IV is managed by the Astrophysical Research Consortium for the Participating Institutions of the SDSS Collaboration including the Brazilian Participation Group, the Carnegie Institution for Science, Carnegie Mellon University, Center for Astrophysics | Harvard \& Smithsonian, the Chilean Participation Group, the French Participation Group, Instituto de Astrof\'isica de Canarias, The Johns Hopkins University, Kavli Institute for the Physics and Mathematics of the Universe (IPMU) / University of Tokyo, the Korean Participation Group, Lawrence Berkeley National Laboratory, Leibniz Institut f\"ur Astrophysik Potsdam (AIP),  Max-Planck-Institut f\"ur Astronomie (MPIA Heidelberg), Max-Planck-Institut f\"ur Astrophysik (MPA Garching), Max-Planck-Institut f\"ur Extraterrestrische Physik (MPE), National Astronomical Observatories of China, New Mexico State University, New York University, University of Notre Dame, Observat\'ario Nacional / MCTI, The Ohio State University, Pennsylvania State University, Shanghai Astronomical Observatory, United Kingdom Participation Group, Universidad Nacional Aut\'onoma de M\'exico, University of Arizona, University of Colorado Boulder, University of Oxford, University of Portsmouth, University of Utah, University of Virginia, University of Washington, University of Wisconsin, Vanderbilt University, and Yale University.

This work has made use of data from the European Space Agency (ESA) mission {\it Gaia} (\url{https://www.cosmos.esa.int/gaia}), processed by the {\it Gaia} Data Processing and Analysis Consortium (DPAC, \url{https://www.cosmos.esa.int/web/gaia/dpac/consortium}). Funding for the DPAC has been provided by national institutions, in particular the institutions participating in the {\it Gaia} Multilateral Agreement.

This research made use of Astropy, a community-developed core Python package for Astronomy (Astropy Collaboration, 2018).
NOIRLab IRAF is distributed by the Community Science and Data Center at NSF NOIRLab, which is managed by the Association of Universities for Research in Astronomy (AURA) under a cooperative agreement with the U.S. National Science Foundation.

\facilities{Keck:I (HIRES Spectrograph), Magellan:Baade (MIKE Spectrograph), Du Pont (APOGEE Spectrograph), Sloan (APOGEE Spectrograph), 2MASS, Gaia}

\software{\href{http://www.astropy.org/}{Astropy}, \href{}{PyRAF}, \href{}{PyKOA}}

\bibliography{Myers}{}
\bibliographystyle{aasjournal}

\clearpage
\appendix

\section{APOGEE Star-by Star Comparison\label{ap:star-by-star}}
\begin{figure*}[hb!]
\begin{center}
    
    \epsscale{0.7}
 	\plotone{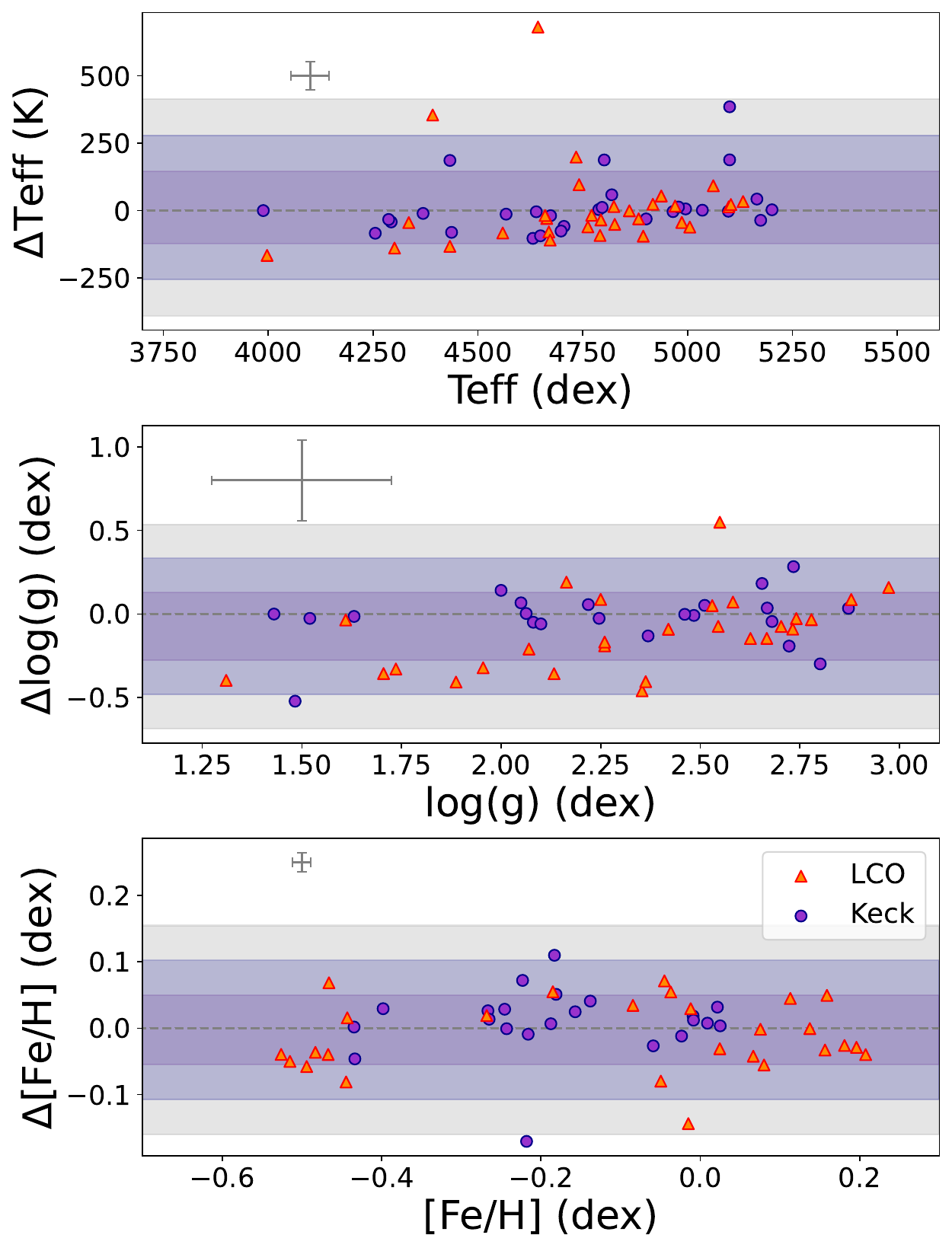}
 	\caption{ \small The difference between the $T_{eff}$ (top), $\log(g)$ (middle), and [Fe/H] (bottom) from this work and APOGEE DR17 as a function of each parameter respectively. Stars which were taken from CG20 and do not have APOGEE data are compared to the $Gaia$ DR2-based $T_{eff}50$ value from CG20. }
 	\label{fig:delta_APOGEE_params}
\end{center}
\end{figure*}

To identify any differences or offsets between our optical data and the APOGEE infrared data, we compare our radial velocity (RV), \teff, \logg, and \feh~values to those in APOGEE DR17. Both the BACCHUS and APOGEE DR17 atmospheric parameters are presented in Table \ref{tab:stellar_params}, 
and a delta plot comparing the difference between \teff, \logg, and \feh~between the two analyses is shown in Figure \ref{fig:delta_APOGEE_params}.
Overall, we see good agreement between the stellar parameters derived here and those in APOGEE DR17, with the vast majority of stars falling within a 1$\sigma$ difference. In \teff, the outliers at $\Delta T > 300$K are all stars which were taken from CG20, and therefore only had \textit{Gaia}-based \teff~measurements.

\noindent \begin{deluxetable*}{lrccccllccrl}
\tabletypesize{\tiny}

\tablecaption{Stellar Parameters (BACCHUS vs DR17) \label{tab:stellar_params}}
\tablehead{\\
 && \multicolumn{4}{c}{BACCHUS} && \multicolumn{4}{c}{APOGEE DR17} \\[1ex]\cline{3-6}  \cline{8-11}\\[-5ex]
    \colhead{Star ID} &
    \colhead{RV} &
    \colhead{\teff\tablenotemark{a}} &
    \colhead{\logg} &
    \colhead{\feh} &
    \colhead{\microturb} &&
    \colhead{\teff} &
    \colhead{\logg} &
    \colhead{\feh} &
    \colhead{RV\tablenotemark{b}} & \\[-6ex]
    \colhead{} &
    \colhead{($km \, s^{-1}$)} & 
    \colhead{(K)} &
    \colhead{(dex)}& 
    \colhead{(dex)} &
    \colhead{($km \, s^{-1}$)} &&
    \colhead{(K)} &
    \colhead{(dex)}& 
    \colhead{(dex)} &
    \colhead{($km \, s^{-1}$)}
}
\startdata
Berkeley 2 $\#$1 & $-76.87 \pm 0.45$ & $ 4293 \pm \phn50$ & $ 1.52 \pm 0.24$ & $-0.24 \pm 0.01$ & $ 1.50 \pm 0.06$ && $ 4335 \pm \phn6$ & $ 1.55 \pm 0.02$ & $-0.24 \pm 0.01$ & $-77.04 \pm 0.05$ \\[0.5ex]
Berkeley 2 $\#$2 & $-76.28 \pm 0.43$ & $ 4901 \pm \phn50$ & $ 2.25 \pm 0.01$ & $-0.22 \pm 0.01$ & $ 1.64 \pm 0.04$ && $ 4932 \pm 10$ & $ 2.27 \pm 0.02$ & $-0.21 \pm 0.01$ & $-76.86 \pm 0.05$ \\[0.5ex]\hline
Berkeley 71 $\#$1 & $-10.78 \pm 0.48$ & $ 5174 \pm \phn50$ & $ 2.80 \pm 0.45$ & $-0.18 \pm 0.01$ & $ 1.25 \pm 0.06$ && $ 5210 \pm 12$ & $ 3.10 \pm 0.02$ & $-0.23 \pm 0.01$ & $-9.54 \pm 0.06$ \\[0.5ex]
Berkeley 71 $\#$5 & $-10.07 \pm 0.33$ & $ 5100 \pm \phn50$ & $ 2.36 \pm 0.03$ & $-0.28 \pm 0.01$ & $ 1.47 \pm 0.06$ && $ 4912 \pm $\nodata & \multicolumn{3}{c}{\em CG20 star not observed by APOGEE} \\[0.5ex]
Berkeley 71 $\#$6 & \phn$-9.10 \pm 0.35$ & $ 5165 \pm \phn50$ & $ 2.67 \pm 0.08$ & $-0.18 \pm 0.01$ & $ 1.59 \pm 0.06$ && $ 5122 \pm 11$ & $ 2.63 \pm 0.02$ & $-0.29 \pm 0.01$ & $-9.33 \pm 0.03$ \\[0.5ex]\hline
Berkeley 75 $\#$1 & $ 125.71 \pm 0.39$ & $ 4369 \pm \phn80$ & $ 1.63 \pm 0.18$ & $-0.43 \pm 0.01$ & $ 1.36 \pm 0.06$ && $ 4379 \pm \phn7$ & $ 1.65 \pm 0.03$ & $-0.44 \pm 0.01$ & $125.01 \pm 0.10$ \\[0.5ex]
Berkeley 75 $\#$1 & $ 125.47 \pm 0.34$ & $ 4335 \pm 122$ & $ 1.61 \pm 0.24$ & $-0.49 \pm 0.01$ & $ 1.48 \pm 0.05$ && $ 4379 \pm \phn7$ & $ 1.65 \pm 0.03$ & $-0.44 \pm 0.01$ & $125.01 \pm 0.10$ \\[0.5ex]
Berkeley 75 $\#$2 & $ 123.86 \pm 0.53$ & $ 4995 \pm \phn50$ & $ 2.73 \pm 0.67$ & $-0.40 \pm 0.01$ & $ 1.44 \pm 0.05$ && $ 4989 \pm 17$ & $ 2.45 \pm 0.04$ & $-0.43 \pm 0.01$ & $124.67 \pm 0.13$ \\[0.5ex]
Berkeley 75 $\#$2 & $ 124.91 \pm 0.32$ & $ 4894 \pm \phn94$ & $ 2.26 \pm 0.33$ & $-0.47 \pm 0.01$ & $ 1.35 \pm 0.06$ && $ 4989 \pm 17$ & $ 2.45 \pm 0.04$ & $-0.43 \pm 0.01$ & $124.67 \pm 0.13$ \\[0.5ex]
Berkeley 75 $\#$4 & $ 123.80 \pm 0.45$ & $ 4965 \pm \phn66$ & $ 2.38 \pm 0.23$ & $-0.47 \pm 0.01$ & $ 1.34 \pm 0.04$ && $ 4969 \pm $\nodata & \multicolumn{3}{c}{\em CG20 star not observed by APOGEE} \\[0.5ex]
Berkeley 75 $\#$4 & $ 126.01 \pm 0.34$ & $ 5061 \pm \phn50$ & $ 2.73 \pm 0.77$ & $-0.40 \pm 0.01$ & $ 1.52 \pm 0.04$ && $ 4969 \pm $\nodata & \multicolumn{3}{c}{\em CG20 star not observed by APOGEE} \\[0.5ex]
Berkeley 75 $\#$5 & $ 125.30 \pm 0.43$ & $ 4978 \pm \phn38$ & $ 2.46 \pm 0.26$ & $-0.41 \pm 0.01$ & $ 1.31 \pm 0.04$ && $ 4965 \pm $\nodata & \multicolumn{3}{c}{\em CG20 star not observed by APOGEE} \\[0.5ex]\hline
 Czernik 20 $\#$1 & $ 30.60 \pm 0.29$ & $ 5035 \pm \phn50$ & $ 2.68 \pm 0.07$ & $-0.14 \pm 0.01$ & $ 1.38 \pm 0.04$ && $ 5034 \pm 14$ & $ 2.73 \pm 0.03$ & $-0.18 \pm 0.01$ & $31.62 \pm 0.01$ \\[0.5ex]
  Czernik 20 $\#$2 & $ 31.45 \pm 0.32$ & $ 5097 \pm \phn50$ & $ 2.87 \pm 0.12$ & $-0.19 \pm 0.01$ & $ 1.20 \pm 0.05$ && $ 5099 \pm 17$ & $ 2.84 \pm 0.03$ & $-0.19 \pm 0.01$ & $31.71 \pm 0.03$ \\[0.5ex]
 Czernik 20 $\#$6 & $ 31.51 \pm 0.27$ & $ 4639 \pm \phn68$ & $ 2.05 \pm 0.04$ & $-0.21 \pm 0.01$ & $ 1.30 \pm 0.05$ && $ 4643 \pm $\nodata & \multicolumn{3}{c}{\em CG20 star not observed by APOGEE} \\[0.5ex]
  Czernik 20 $\#$7 & $ 31.75 \pm 0.27$ & $ 4801 \pm \phn50$ & $ 2.56 \pm 0.03$ & $-0.15 \pm 0.02$ & $ 1.39 \pm 0.06$ && $ 4613 \pm $\nodata & \multicolumn{3}{c}{\em CG20 star not observed by APOGEE} \\[0.5ex]\hline
 Czernik 30 $\#$2 & $ 81.14 \pm 0.30$ & $ 4567 \pm 109$ & $ 2.06 \pm 0.28$ & $-0.43 \pm 0.01$ & $ 1.39 \pm 0.05$ && $ 4580 \pm 12$ & $ 2.06 \pm 0.03$ & $-0.39 \pm 0.01$ & $81.68 \pm 0.04$ \\[0.5ex]\hline
 ESO 518 03 $\#$1 & $ 21.51 \pm 0.13$ & $ 4734 \pm \phn61$ & $ 2.33 \pm 0.01$ & $+0.04 \pm 0.01$ & $ 1.41 \pm 0.05$ && $ 4535 \pm $\nodata & \multicolumn{3}{c}{\em CG20 star not observed by APOGEE} \\[0.5ex]
 ESO 518 03 $\#$2 & $ 22.36 \pm 0.17$ & $ 3997 \pm 196$ & $ 1.31 \pm 0.62$ & $-0.05 \pm 0.02$ & $ 1.27 \pm 0.09$ && $ 4163 \pm \phn5$ & $ 1.71 \pm 0.02$ & $+0.03 \pm 0.01$ & $22.30 \pm 0.05$ \\[0.5ex]
 ESO 518 03 $\#$4 & $ 22.22 \pm 0.11$ & $ 4824 \pm \phn50$ & $ 2.97 \pm 0.10$ & $+0.08 \pm 0.01$ & $ 1.21 \pm 0.05$ && $ 4809 \pm \phn8$ & $ 2.81 \pm 0.02$ & $+0.08 \pm 0.01$ & $22.71 \pm 0.07$ \\[0.5ex]
 ESO 518 03 $\#$5 & $ 21.85 \pm 0.12$ & $ 4771 \pm 107$ & $ 2.55 \pm 0.25$ & $+0.07 \pm 0.01$ & $ 1.31 \pm 0.05$ && $ 4788 \pm \phn9$ & $ 2.62 \pm 0.02$ & $+0.11 \pm 0.01$ & $22.13 \pm 0.00$ \\[0.5ex]\hline
 Haffner 4 $\#$1 & $ 57.30 \pm 0.38$ & $ 5201 \pm \phn50$ & $ 2.72 \pm 0.05$ & $-0.16 \pm 0.01$ & $ 1.54 \pm 0.05$ && $ 5198 \pm 18$ & $ 2.92 \pm 0.03$ & $-0.18 \pm 0.01$ & $59.65 \pm 0.21$ \\[0.5ex]
 Haffner 4 $\#$2 & $ 58.65 \pm 0.27$ & $ 4433 \pm \phn50$ & $ 2.51 \pm 0.05$ & $-0.21 \pm 0.02$ & $ 1.59 \pm 0.12$ && $ 4247 \pm $\nodata & \multicolumn{3}{c}{\em CG20 star not observed by APOGEE} \\[0.5ex]
 Haffner 4 $\#$3 & $ 58.93 \pm 0.35$ & $ 5100 \pm \phn50$ & $ 2.52 \pm 0.10$ & $-0.13 \pm 0.01$ & $ 1.49 \pm 0.05$ && $ 4715 \pm $\nodata & \multicolumn{3}{c}{\em CG20 star not observed by APOGEE} \\[0.5ex]\hline
 NGC 1798 $\#$1 & $ nan \pm nan$ & $ 4437 \pm 151$ & $ 2.00 \pm 0.11$ & $-0.22 \pm 0.01$ & $ 1.32 \pm 0.07$ && $ 4518 \pm\phn8$ & $ 1.86 \pm 0.03$ & $-0.30 \pm 0.01$ & $4.06 \pm 0.03$ \\[0.5ex]
 NGC 1798 $\#$2 & $ 2.78 \pm 0.25$ & $ 4631 \pm \phn 50$ & $ 2.08 \pm 0.06$ & $-0.27 \pm 0.01$ & $ 1.40 \pm 0.05$ && $ 4734 \pm 10$ & $ 2.13 \pm 0.03$ & $-0.29 \pm 0.01$ & $2.69 \pm 0.04$ \\[0.5ex]
 NGC 1798 $\#$3 & $ 2.48 \pm 0.26$ & $ 4649 \pm 109$ & $ 2.10 \pm 0.14$ & $-0.25 \pm 0.01$ & $ 1.31 \pm 0.05$ && $ 4742 \pm\phn8$ & $ 2.16 \pm 0.02$ & $-0.27 \pm 0.01$ & $2.69 \pm 0.05$ \\[0.5ex]
 NGC 1798 $\#$4 & $ 1.65 \pm 0.28$ & $ 4673 \pm 126$ & $ 2.22 \pm 0.52$ & $-0.26 \pm 0.01$ & $ 1.29 \pm 0.05$ && $ 4692 \pm\phn9$ & $ 2.16 \pm 0.03$ & $-0.28 \pm 0.01$ & $2.07 \pm 0.03$ \\[0.5ex]\hline
 NGC 2243 $\#$1 & $ 60.20 \pm 0.17$ & $ 4793 \pm \phn50$ & $ 2.58 \pm 0.59$ & $-0.48 \pm 0.01$ & $ 1.40 \pm 0.04$ && $ 4828 \pm 15$ & $ 2.51 \pm 0.04$ & $-0.45 \pm 0.01$ & $60.29 \pm 0.01$ \\[0.5ex]
 NGC 2243 $\#$2 & $ 60.51 \pm 0.14$ & $ 4826 \pm \phn59$ & $ 2.42 \pm 0.27$ & $-0.53 \pm 0.01$ & $ 1.34 \pm 0.05$ && $ 4877 \pm 13$ & $ 2.51 \pm 0.03$ & $-0.49 \pm 0.01$ & $60.21 \pm 0.08$ \\[0.5ex]
 NGC 2243 $\#$3 & $ 60.34 \pm 0.15$ & $ 4791 \pm \phn79$ & $ 2.07 \pm 0.22$ & $-0.51 \pm 0.01$ & $ 1.41 \pm 0.04$ && $ 4883 \pm 11$ & $ 2.28 \pm 0.03$ & $-0.47 \pm 0.01$ & $60.21 \pm 0.04$ \\[0.5ex]
 NGC 2243 $\#$4 & $ 60.34 \pm 0.18$ & $ 4664 \pm \phn50$ & $ 2.16 \pm 0.58$ & $-0.47 \pm 0.01$ & $ 1.52 \pm 0.04$ && $ 4693 \pm \phn9$ & $ 1.97 \pm 0.03$ & $-0.53 \pm 0.01$ & $60.47 \pm 0.04$ \\[0.5ex]
 NGC 2243 $\#$5 & $ 60.31 \pm 0.14$ & $ 4986 \pm \phn50$ & $ 2.53 \pm 0.60$ & $-0.44 \pm 0.01$ & $ 1.47 \pm 0.04$ && $ 5030 \pm 15$ & $ 2.48 \pm 0.03$ & $-0.46 \pm 0.01$ & $60.24 \pm 0.09$ \\[0.5ex]\hline
 NGC 2447 $\#$1 & $ 24.56 \pm 0.13$ & $ 5098 \pm \phn150$ & $ 2.63 \pm 0.03$ & $-0.08 \pm 0.01$ & $ 1.41 \pm 0.04$ && $ 5084 \pm 10$ & $ 2.77 \pm 0.02$ & $-0.12 \pm 0.01$ & $24.22 \pm 0.12$ \\[0.5ex]
 NGC 2447 $\#$2 & $ 22.75 \pm 0.12$ & $ 5103 \pm \phn50$ & $ 2.74 \pm 0.03$ & $-0.05 \pm 0.01$ & $ 1.40 \pm 0.04$ && $ 5080 \pm 10$ & $ 2.77 \pm 0.02$ & $-0.12 \pm 0.01$ & $22.63 \pm 0.15$ \\[0.5ex]
 NGC 2447 $\#$3 & $ 23.48 \pm 0.10$ & $ 5132 \pm \phn50$ & $ 2.70 \pm 0.05$ & $-0.04 \pm 0.01$ & $ 1.43 \pm 0.04$ && $ 5098 \pm 10$ & $ 2.78 \pm 0.02$ & $-0.09 \pm 0.01$ & $23.50 \pm 0.06$ \\[0.5ex]\hline
 NGC 2632 $\#$1 & $ 34.98 \pm 0.19$ & $ 5190 \pm \phn67$ & $ 4.66 \pm 0.47$ & $+0.14 \pm 0.01$ & $ 1.21 \pm 0.04$ && $ 5165 \pm 15$ & $ 4.53 \pm 0.02$ & $+0.18 \pm 0.01$ & $35.05 \pm 0.04$ \\[0.5ex]
 NGC 2632 $\#$2 & $ 34.77 \pm 0.36$ & $ 4498 \pm 123$ & $ 4.27 \pm 1.29$ & $-0.12 \pm 0.03$ & $ 1.10 \pm 0.08$ && $ 4335 \pm 10$ & $ 4.58 \pm 0.02$ & $+0.11 \pm 0.01$ & $35.03 \pm 0.11$ \\[0.5ex]
 NGC 2632 $\#$3 & $ 34.96 \pm 0.17$ & $ 5847 \pm \phn50$ & $ 4.58 \pm 0.22$ & $+0.15 \pm 0.01$ & $ 1.24 \pm 0.04$ && $ 5712 \pm 23$ & $ 4.40 \pm 0.02$ & $+0.17 \pm 0.00$ & $35.25 \pm 0.01$ \\[0.5ex]\hline
 NGC 2682 $\#$1 & $ 34.57 \pm 0.47$ & $ 4287 \pm 139$ & $ 1.48 \pm 0.14$ & $-0.06 \pm 0.02$ & $ 1.25 \pm 0.07$ && $ 4320 \pm \phn6$ & $ 2.00 \pm 0.02$ & $-0.03 \pm 0.01$ & $34.07 \pm 0.12$ \\[0.5ex]
 NGC 2682 $\#$2 & $ 34.07 \pm 0.48$ & $ 4255 \pm 132$ & $ 2.05 \pm 0.47$ & $-0.01 \pm 0.02$ & $ 1.23 \pm 0.08$ && $ 4339 \pm \phn6$ & $ 1.98 \pm 0.02$ & $-0.03 \pm 0.01$ & $33.53 \pm 0.00$ \\[0.5ex]
 NGC 2682 $\#$3 & $ 33.96 \pm 0.29$ & $ 4705 \pm \phn 50$ & $ 2.48 \pm 0.04$ & $+0.01 \pm 0.01$ & $ 1.29 \pm 0.05$ && $ 4764 \pm 11$ & $ 2.49 \pm 0.03$ & $+0.00 \pm 0.01$ & $33.77 \pm 0.00$ \\[0.5ex]
 NGC 2682 $\#$4 & $ 35.43 \pm 0.81$ & $ 3988 \pm \phn76$ & $ 1.43 \pm 1.05$ & $-0.22 \pm 0.02$ & $ 1.54 \pm 0.14$ && $ 3988 \pm \phn5$ & $ 1.43 \pm 0.02$ & $-0.05 \pm 0.01$ & $32.61 \pm 0.00$ \\[0.5ex]
 NGC 2682 $\#$5 & $ 34.46 \pm 0.31$ & $ 4819 \pm 156$ & $ 2.51 \pm 0.74$ & $+0.03 \pm 0.01$ & $ 1.32 \pm 0.07$ && $ 4760 \pm \phn8$ & $ 2.46 \pm 0.02$ & $+0.02 \pm 0.01$ & $34.05 \pm 0.05$ \\[0.5ex]
 NGC 2682 $\#$6 & $ 36.02 \pm 0.28$ & $ 4788 \pm \phn98$ & $ 2.65 \pm 0.63$ & $+0.02 \pm 0.01$ & $ 1.32 \pm 0.06$ && $ 4784 \pm \phn8$ & $ 2.47 \pm 0.02$ & $-0.01 \pm 0.01$ & $34.18 \pm 0.01$ \\[0.5ex]
 NGC 2682 $\#$7 & $ 34.66 \pm 0.73$ & $ 4796 \pm 114$ & $ 2.46 \pm 0.38$ & $-0.01 \pm 0.01$ & $ 1.33 \pm 0.05$ && $ 4784 \pm \phn8$ & $ 2.46 \pm 0.02$ & $-0.02 \pm 0.01$ & $34.59 \pm 0.03$ \\[0.5ex]
 NGC 2682 $\#$8 & $ 34.95 \pm 0.29$ & $ 4698 \pm \phn50$ & $ 2.37 \pm 0.25$ & $-0.02 \pm 0.01$ & $ 1.25 \pm 0.05$ && $ 4774 \pm \phn8$ & $ 2.50 \pm 0.02$ & $-0.01 \pm 0.01$ & $34.47 \pm 0.02$ \\[0.5ex]\hline
 NGC 4337 $\#$1 & $-17.12 \pm 0.13$ & $ 4917 \pm 184$ & $ 2.78 \pm 0.84$ & $+0.20 \pm 0.01$ & $ 1.33 \pm 0.06$ && $ 4893 \pm \phn8$ & $ 2.81 \pm 0.02$ & $+0.22 \pm 0.01$ & $-17.06 \pm 0.07$ \\[0.5ex]
 NGC 4337 $\#$3 & $-18.43 \pm 0.11$ & $ 4937 \pm 151$ & $ 2.88 \pm 0.82$ & $+0.18 \pm 0.01$ & $ 1.36 \pm 0.06$ && $ 4883 \pm \phn8$ & $ 2.79 \pm 0.02$ & $+0.21 \pm 0.01$ & $-17.77 \pm 0.07$ \\[0.5ex]
 NGC 4337 $\#$4 & $-17.61 \pm 0.12$ & $ 4861 \pm 178$ & $ 2.67 \pm 0.76$ & $+0.21 \pm 0.01$ & $ 1.33 \pm 0.06$ && $ 4862 \pm \phn9$ & $ 2.81 \pm 0.02$ & $+0.25 \pm 0.01$ & $-17.81 \pm 0.08$ \\[0.5ex]
 NGC 4337 $\#$5 & $-18.19 \pm 0.12$ & $ 4883 \pm 120$ & $ 2.73 \pm 0.24$ & $+0.16 \pm 0.01$ & $ 1.32 \pm 0.06$ && $ 4914 \pm 10$ & $ 2.82 \pm 0.02$ & $+0.19 \pm 0.01$ & $-18.68 \pm 0.33$ \\[0.5ex]\hline
 NGC 6705 $\#$1 & $ 36.06 \pm 0.15$ & $ 4762 \pm \phn50$ & $ 2.35 \pm 0.15$ & $+0.16 \pm 0.01$ & $ 1.60 \pm 0.11$ && $ 4822 \pm \phn8$ & $ 2.81 \pm 0.02$ & $+0.11 \pm 0.01$ & $36.31 \pm 0.04$ \\[0.5ex]
 NGC 6705 $\#$2 & $ 35.93 \pm 0.14$ & $ 4669 \pm 111$ & $ 2.26 \pm 0.19$ & $+0.14 \pm 0.02$ & $ 1.68 \pm 0.10$ && $ 4748 \pm \phn8$ & $ 2.43 \pm 0.02$ & $+0.14 \pm 0.01$ & $35.44 \pm 0.02$ \\[0.5ex]\hline
 Ruprecht 82 $\#$2 & $ 2.33 \pm 0.12$ & $ 5005 \pm \phn50$ & $ 2.36 \pm 0.09$ & $-0.01 \pm 0.01$ & $ 1.46 \pm 0.05$ && $ 5066 \pm 10$ & $ 2.77 \pm 0.02$ & $-0.04 \pm 0.01$ & $1.99 \pm 0.06$ \\[0.5ex]\hline
 Ruprecht 85 $\#$1 & $ 21.87 \pm 0.16$ & $ 4672 \pm \phn50$ & $ 1.71 \pm 0.05$ & $-0.27 \pm 0.01$ & $ 1.44 \pm 0.08$ && $ 4781 \pm \phn9$ & $ 2.06 \pm 0.02$ & $-0.29 \pm 0.01$ & $21.80 \pm 0.06$ \\[0.5ex]
 Ruprecht 85 $\#$2 & $ 22.02 \pm 0.24$ & $ 4970 \pm 161$ & $ 2.13 \pm 0.48$ & $-0.19 \pm 0.01$ & $ 1.57 \pm 0.08$ && $ 4953 \pm 10$ & $ 2.49 \pm 0.02$ & $-0.24 \pm 0.01$ & $21.85 \pm 0.05$ \\[0.5ex]
 Ruprecht 85 $\#$3 & $ 24.47 \pm 0.14$ & $ 4643 \pm \phn50$ & $ 2.13 \pm 0.21$ & $-0.11 \pm 0.02$ & $ 1.74 \pm 0.08$ && $ 3962 \pm $\nodata & \multicolumn{3}{c}{\em CG20 star not observed by APOGEE} \\[0.5ex]
 Ruprecht 85 $\#$4 & $ 17.60 \pm 0.14$ & $ 4392 \pm 228$ & $ 1.42 \pm 0.52$ & $-0.25 \pm 0.01$ & $ 1.51 \pm 0.09$ && $ 4037 \pm $\nodata & \multicolumn{3}{c}{\em CG20 star not observed by APOGEE} \\[0.5ex]\hline
 SAI 116 $\#$1  & $-14.53 \pm 0.17$ & $ 4660 \pm \phn88$ & $ 1.89 \pm 0.05$ & $+0.08 \pm 0.02$ & $ 1.74 \pm 0.09$ && $ 4678 \pm \phn7$ & $ 2.29 \pm 0.02$ & $+0.14 \pm 0.01$ & $-14.25 \pm 0.05$ \\
 SAI 116 $\#$2  & $-13.94 \pm 0.11$ & $ 4559 \pm \phn52$ & $ 1.96 \pm 0.18$ & $-0.01 \pm 0.01$ & $ 1.72 \pm 0.10$ && $ 4642 \pm \phn7$ & $ 2.28 \pm 0.02$ & $+0.13 \pm 0.01$ & $-13.95 \pm 0.08$ \\[0.5ex]\hline
 Tombaugh 2 $\#$1 & $ 122.17 \pm 0.26$ & $ 4741 \pm \phn54?$ & $ 2.55 \pm 0.39$ & $-0.44 \pm 0.02$ & $ 1.50 \pm 0.06$ && $ 4645 \pm 12$ & $ 2.00 \pm 0.03$ & $-0.36 \pm 0.01$ & $121.41 \pm 0.14$ \\[0.5ex]\hline
Trumpler 20 $\#$4 & $-40.00 \pm 0.16$ & $ 4301 \pm \phn 50$ & $ 1.74 \pm 0.13$ & $+0.02 \pm 0.01$ & $ 1.23 \pm 0.07$ && $ 4440 \pm \phn6$ & $ 2.07 \pm 0.02$ & $+0.06 \pm 0.01$ & $-40.79 \pm 0.02$ \\[0.5ex]
Trumpler 20 $\#$5 & $-36.87 \pm 0.13$ & $ 4433 \pm \phn 50$ & $ 2.25 \pm 0.26$ & $+0.11 \pm 0.01$ & $ 1.30 \pm 0.07$ && $ 4566 \pm \phn7$ & $ 2.16 \pm 0.02$ & $+0.07 \pm 0.01$ & $-39.44 \pm 0.37$ \\[0.5ex]\hline
\enddata
\tablenotetext{a}{\teff~errors include a conservative floor of 50K.}\vskip-0.1in
\tablenotetext{b}{The error reported comes from the \texttt{VSCATTER}}\vskip-0.1in
\tablenotetext{c}{Data taken at Las Campanas Observatory}\vskip-0.1in
\tablenotetext{d}{Data taken at the W. M. Keck Observatory}
\end{deluxetable*}

\clearpage
\section{Gradient Comparisons with the Literature }
Here, we compare the neutron-capture abundance gradients presented in Section \ref{sec:results} with those in the literature.  We also recalculate our gradients relative to Fe for consistency between surveys.  
For these comparisons, we refer to (1) \citet{magrini2023-gradients-withESO-clusters} which uses the Gaia-ESO Survey (GES), (2) The Open Clusters Chemical Abundances from Spanish Observatories Survey\citep[OCCASO;][]{casamiquela2016-RVs, casamiquela2017-feh, casamiquela2019-fehgradient, casamiquela2022-ocmembership-in51ocs, occasoV}, (3) \citet{jacobson-oc-srprocess}, and (4) \citet{Otto-occam}. 
We include a table of neutron-capture gradients measured in the literature for reference (Table \ref{tab:litgrads}).

{In the first-peak $s-$process abundances, we find that we agree within the $1\sigma$ error. Particularly, we find strong agreement in our slope for \yfe~($+0.012\pm 0.019$ dex kpc$^{-1}$) compared to the GES gradient ($0.012\pm 0.003$ dex kpc$^{-1}$), and relatively poor agreement with the OCCASO gradient ($-0.005\pm0.011$ dex kpc$^{-1}$), where both OCCASO and our measurements are consistent with a flat slope. We also find relatively poor agreement between the \zrfe gradient measured here ($+0.020\pm 0.019$ dex kpc$^{-1}$) and in GES ($0.005\pm 0.003$ dex kpc$^{-1}$), although the OCCASO gradient ($0.009\pm 0.015$ dex kpc$^{-1}$) is more similar to the one measured by GES the gradient measured in \citet{jacobson-oc-srprocess} is more similar to ours ($0.029\pm 0.005$ dex kpc$^{-1}$). 
On a cluster-by-cluster basis, our \zrfe~measurements all fall within $\pm0.25$ dex of the GES measurements, however, they do have a significantly enhanced cluster in their sample that is not present in this work (NGC 6791). This cluster is located very near to the solar radius ($7.94$ kpc) but has much more enhanced \zrfe~abundance ($0.36$ dex), which could work to flatten their abundance curve. }

{In the second-peak $s-$process abundances, we find agreement within the $1\sigma$ errors for the \bafe~and \lafe~slopes between all studies we compare against. In particular, we measure a slope of $0.032\pm0.019$ dex kpc$^{-1}$ for \bafe~and $0.038\pm0.019$ dex kpc$^{-1}$ for \lafe.  GES reported values of $0.019\pm0.006$ dex kpc$^{-1}$ and $0.021\pm0.004$ dex kpc$^{-1}$ for \bafe~and \lafe, respectively; \citet{jacobson-oc-srprocess} measures gradients of $0.039\pm 0.006$ dex kpc$^{-1}$ for \bafe~and $0.036\pm 0.003$ dex kpc$^{-1}$ for \lafe; and OCCASO measures gradients of $0.026\pm 0.018$ dex kpc$^{-1}$ for \bafe. }
{However, for the \cefe~abundances, our slope ($0.059\pm0.019$ dex kpc$^{-1}$) is much steeper than the GES and OCCASO \cefe~gradients ($0.014\pm0.003$ dex kpc$^{-1}$ and $0.001\pm 0.013$ dex kpc$^{-1}$). We do see a radial trend in the difference between our abundances and those measured in GES, that is, we find a larger abundance for clusters that are at larger radii and a smaller abundance for  clusters at inner radii. This would have the affect of steepening the radial gradient for \cefe.   We do find good agreement with the gradient measured with the \cefe~gradient from OCCAM ($0.087\pm 0.007$ dex kpc$^{-1}$).}

{Finally, for the two $r-$process elements we have in common, \ndfe~and \eufe, we find good agreement within the standard deviation. In \ndfe~we measure a slope of: $0.057\pm0.018$ dex kpc$^{-1}$, GES measures a slope of: $0.045\pm0.006$ dex kpc$^{-1}$, OCCASO measures a slope of: $0.032\pm 0.015$ dex kpc$^{-1}$. For \eufe~we find a slope of: $0.029\pm0.020$ dex kpc$^{-1}$, which is similar to the gradient measured in GES: $0.030\pm0.005$ dex kpc$^{-1}$, and that measured in \citealt{jacobson-oc-srprocess}: $0.047\pm 0.003$ dex kpc$^{-1}$. }

\begin{deluxetable*}{lcccc}
\tablecaption{Literature Linear [X/Fe] Abundance Gradients  \label{tab:litgrads}}
\tablehead{
\colhead{Reference} &
\colhead{Element} &
\colhead{Gradient} &
\colhead{Radius Range} &
\colhead{N clusters} \\[-2ex]
\colhead{} &
\colhead{} &
\colhead{(dex kpc$^{-1}$)} &
\colhead{(kpc)} &
\colhead{}}



\startdata
This Work & \yfe & $+0.012\pm0.019$ & 6--17 & 18 \\
Gaia-ESO$^a$ & \yfe & $ +0.012\pm0.003 $ & 6--21 & 62 \\
OCCASO$^b$ & \yfe & $ -0.005\pm0.011 $ & 6--12 & 36 \\\hline
This Work & \zrfe & $+0.020\pm0.019$ & 6--17 & 18 \\
Gaia-ESO$^a$ & \zrfe & $ +0.005\pm0.003 $ & 6--17 & 56 \\
OCCASO$^b$ & \zrfe & $ +0.009\pm0.015 $ & 6--12 & 35 \\
Jacobson$^c$ & \zrfe & $ +0.029\pm0.005 $ & 9--15 & 19 \\\hline
This Work & \bafe & $+0.032\pm0.019$ & 6--17 & 18 \\
Gaia-ESO$^a$ & \bafe & $ +0.019\pm0.006 $ & 6--21 & 60 \\
OCCASO$^b$ & \bafe & $ +0.026\pm0.018 $ & 6--12 & 36 \\
Jacobson$^c$ & \bafe & $ +0.039\pm0.006 $ & 9--15 & 19 \\\hline
This Work & \lafe & $+0.038\pm0.019$ & 6--17 & 18 \\
Gaia-ESO$^a$ & \lafe & $ +0.021\pm0.004 $ & 6--17 & 52\\
Jacobson$^c$ & \lafe & $ +0.036\pm0.003 $ & 9--15 & 19 \\\hline
This Work & \cefe & $+0.059\pm0.019$ & 6--17 & 18 \\
Gaia-ESO$^a$ & \cefe & $ +0.014\pm0.003 $ & 6--17 & 56 \\
OCCASO$^b$ & \cefe & $ +0.001\pm0.013 $ & 6--12 & 36 \\
OCCAM VIII$^d$ & \cefe & $+0.087\pm0.007$ & 6--21 & 142\\\hline
This Work & \ndfe & $+0.057\pm0.018$ & 6--17 & 18 \\
Gaia-ESO$^a$ & \ndfe & $ +0.045\pm0.006 $ & 6--21 & 61 \\
OCCASO$^b$ & \ndfe & $ +0.032\pm0.015 $ & 6--12 & 36 \\\hline
This Work & \eufe & $+0.029\pm0.020$ & 6--17 & 8 \\
Gaia-ESO$^a$ & \eufe & $ +0.030\pm0.005 $ & 6--21 & 59 \\
Jacobson$^c$ & \eufe & $ +0.047\pm0.003 $ & 9--15 & 19 \\
\enddata
\tablenotetext{a}{\citet{magrini2023-gradients-withESO-clusters}.}\vskip-0.1in
\tablenotetext{b}{\citet{occasoV}.}\vskip-0.1in
\tablenotetext{c}{\citet{jacobson-oc-srprocess}.}\vskip-0.1in
\tablenotetext{d}{\citet{Otto-occam}.}
\end{deluxetable*}

\end{document}